\documentclass[aps,pre, reprint, groupedaddress,superscriptaddress, amsmath,amssymb]{revtex4-1}

\usepackage[mathlines]{lineno}% Enable numbering of text and display math
\usepackage{color}
\usepackage{float}

\usepackage{graphicx}% Include figure files

\usepackage{wrapfig}
\usepackage{wasysym}

\usepackage{adjustbox}
\usepackage{ragged2e}
\usepackage[svgnames,table]{xcolor}
\usepackage{array}
\usepackage{changepage}
\usepackage{setspace}
\usepackage{hhline}
\usepackage{tabto}
\usepackage{multirow}
\usepackage{makecell}

\usepackage{dcolumn}% Align table columns on decimal point
\usepackage{bm}% bold math

\usepackage{flowchart}

\begin{document}

%\title{Effect of the Brine Composition on the Dissolution of CO$_2$ into Brine: Flux Growth; Quasi-Steady and Shut-Down Regimes}

\title{Convective dissolution of carbon dioxide in deep saline aquifers:\\Insights from engineering a  high-pressure porous Hele-Shaw cell}

\author{Saeed Mahmoodpour}
\affiliation{Institute of Petroleum Engineering, College  of  Engineering, University of Tehran, Iran}

\author{Behzad Rostami}
\email[]{brostami@ut.ac.ir}
\affiliation{Institute of Petroleum Engineering, College of  Engineering, University of Tehran, Iran}

\author{Mohamad Reza Soltanian}
\affiliation{Department of Geology, University of Cincinnati,  Cincinnati, Ohio 45221, USA}
\affiliation{Department of Chemical and Environmental Engineering, University of Cincinnati,  Cincinnati, Ohio 45221, USA}

\author{Mohammad Amin Amooie}
\email[]{amooie@mit.edu}
%\affiliation{School of Earth Sciences, The Ohio State University, Columbus, Ohio, USA}
\affiliation{Department of Chemical Engineering, Massachusetts Institute of Technology, Cambridge, Massachusetts 02139, USA}

\newcommand{\ext}{pdf}

\date{\today}

%overlying brine in the porous region as well as

\begin{abstract}
We present the first  experiments of   dissolution-driven convection  of carbon dioxide (CO$_2$) in a confined brine-saturated porous medium at high pressures. We designed a novel Hele-Shaw cell  that allows for both visual and quantitative analyses, and   address the effects of free-phase CO$_2$ and  brine composition on convective dissolution.  The visual examination of the gas volume combined with  the measurement of pressure, which both evolve with dissolution, enable us to yield  insights into the dynamics of convection in conditions that more closely reflect the geologic conditions. We find and analyze different dissolution events, including diffusive, early and late convection, and shut-down regimes. Our experiments reveal that in intermediate regime, a so-called ``quasi-steady'' state actually never happens. Dissolution flux continuously decreases in this regime, which  is due to a negative feedback loop: the rapid reduction of pressure following convective dissolution, in turn, decreases the solubility of CO$_2$ at the gas-brine interface and thus the instability strength. We introduce a new scaling factor that  not only compensates the flux reduction but also the nonlinearities that arise from different salt types. We present robust scaling relations for  the compensated flux and for the transition times between consecutive regimes in  systems with NaCl (Ra $\sim$ 3271--4841) and  NaCl+CaCl$_2$ mixtures (Ra $\sim$ 2919--4283). We also find that  NaCl+CaCl$_2$ mixtures enjoy a longer intermediate period before the shut-down of dissolution, but with a lower dissolution flux, as compared to NaCl brines. The results provide a new perspective into
how the presence of two separate phases in a closed system as well as different salt types may affect the predictive powers of our experiments and  models  for both the
short- and long-term dynamics of convective dissolution  in porous media.

\end{abstract}

\pacs{}

\maketitle

\section{\label{sec1}Introduction}

Natural convection emerges when a fluid with higher density  overlays a lower density fluid. Convection  can significantly enhance the transport
of mass, heat, and energy, and is ubiquitous
 in many natural and industrial areas including
weather systems  \citep{doswell1987distinction, hodges1997distribution}, plate
tectonics  \citep{elder1967steady, gurnis1988large}, oceanic currents 
\citep{kampf1998shallow},   sea water and
groundwater aquifers \citep{van2009natural}, free air cooling  \citep{fedorov1997turbulent}, electronic
devices  \citep{rolin2014mass, hu2018binary, ordonez2016transistorlike},
solar ponds  \citep{suarez2010fully}, and microfluidic systems 
\citep{zehavi2016competition}. In the context of porous media, natural convection has
applications in geothermal energy production \citep{bejan1985heat}, heat exchangers
\citep{cheng1977free}, sand separation from oil \citep{mansour2015numerical}, and many others. The convection in porous media has recently
received renewed attention because of its importance in carbon dioxide (CO$_2$) sequestration in the underground
formations as  one of the most promising options to
stabilize atmospheric CO$_2$ concentrations and hence mitigate
the global climate change \citep{coninck2005ipcc}. Deep saline aquifers have been recognized as the primary target among storage repositories beneath the Earth's surface, mainly owing to  their favorable chemistry, porosity,
temperature, pressure, huge capacity, and wide distribution all over the world \citep{keller2014potential}. Natural
convection controls the dissolution of injected CO$_2$ into underlying brine, which is a key mechanism for
the permanent and  efficient trapping of CO$_2$. Our goal here is to study the dynamics of convective CO$_2$ dissolution in brine, which not only gives insights into the short- and long-term fate of  CO$_2$ injected into the subsurface but may also contribute to the further understanding of   convection dynamics in other fields.\\

%dissolution regimes in this regard, and interaction between convective fingers are notoriously difficult. Because, there is need for coupling the qualitative (image) and quantitative (such as pressure) data in a high pressure set-up. 

During the geological sequestration of CO$_2$ in deep brine-bearing formations (i.e., saline aquifers), buoyant CO$_2$ rises upward until it is confined by an impermeable caprock while  spreading laterally beneath.  This \textit{structurally} trapped CO$_2$ is, however, susceptible to leak back to the surface due to the existence of small fractures or faults in the seal. Moreover, in the injection stage the caprock integrity may be compromised by the reservoir overpressurization, which would induce new fractures or cause slip along pre-existing faults in the seal \citep{szulczewski2013carbon, soltanian2018impacts}. \\

CO$_2$ dissolution in brine (or groundwater) is an important trapping mechanism towards permanent storage of CO$_2$,   which would reduce the risk of leakage from imperfect or compromised caprocks \citep{javaheri2010linear, szulczewski2013carbon, martinez2016two, soltanian2016simulating, newell2018experimental, shi2018measurement, dai2018heterogeneity}. The diffusion of CO$_2$\textsubscript{ }molecules into brine initiates the dissolution process \citep{soltanian2016critical, soltanian2017dissolution}, which results in a diffusive boundary layer that is more dense than the underlying formation brine and hence prone to density-driven instabilities \citep{soltanian2017dissolution}. Beyond a critical thickness of this boundary layer, fingering instabilities will form. As these instabilities grow, they migrate toward the formation bottom through convection while carrying further dissolved CO$_2$ away from the seal. The underlying lighter fluid at lower CO$_2$ concentrations rises upward at the same time, thereby sharpening the CO$_2$ concentration gradient at the  gas-brine interface that accelerates the dissolution rate \citep{hassanzadeh2005modelling, hassanzadeh2007scaling, farajzadeh2007numerical, emami2015convective}. An estimation of the dissolution flux helps to constrain the amount of CO$_2$ that will remain in solution in the subsurface and the amount that is prone to escape.\\
 
The dissolution and  mixing of CO$_2$ has been well studied in idealized systems of porous convection. The \textit{two-phase} fluid system that forms following the injection of CO$_2$, where free-phase (gaseous) CO$_2$ overlays the brine-saturated layer with a dynamic and temporally evolving interface in between, is typically simplified into a \textit{one-phase} system through one of the following approaches. Analog fluid systems are used as an alternative, where the two-phase CO$_2$-brine system is replaced with a single-phase two-layer system composed of water and suitable fluid (e.g., MEG or PPG) that is miscible with water. As such, the analog fluid systems result in non-monotonic density profile for mixture, and cannot represent the partial miscibility \citep{amooie2017hydrothermodynamic, amooie2017mixing}, density and viscosity profiles,  instability strength, and thus the underlying dynamics of actual CO$_2$-brine systems  \citep{backhaus2011convective, hidalgo2012scaling, raad2015onset}. The two-phase system models can be also simplified by  including only the porous layer below the gas-brine interface and representing this interface by  (\textit{i}) a top boundary fixed at CO$_2$ saturation\citep[e.g.,][]{pau2010high, slim2013dissolution, de2017dissolution} or  (\textit{ii}) a top boundary of constant flux of CO$_2$ at low rate such that CO$_2$ goes into solution immediately \citep{soltanian2016critical, soltanian2017dissolution, amooie2018solutal}. \\

In these systems, dissolution process after the \textit{onset of convection} can be summarized as follows: First, the fingers grow independently and descend to the bottom of  aquifer with small lateral interactions \citep{slim2013dissolution}. However, as time goes on fingers start to merge to their neighboring fingers and create stronger fingers. At this time,  the dissolution flux (or mixing rate) increases with time in a \textit{flux-growth } (or dissipation-growth) regime \citep{slim2014solutal, wen2018dynamics}. After the convective regime has well developed, the \textit{quasi-steady state }regime starts \citep{wen2018dynamics}, during which the  convective flow brings the fresh brine to the gas-brine interface and the process continues in a \textit{quasi-steady }regime until the  brine brought to the gas-brine interface begins to contain dissolved CO$_2$.  The latter can happen some time after the descending fingers reach the bottom.  The dissolution process then shifts to the \textit{shut-down }regime and continues to reach the maximum dissolution capacity of system at the given pressure, temperature, and salinity \citep{riaz2014carbon}. For completeness, we gather and summarize the scaling results from previous numerical and experimental studies  for these regimes in \textbf{Table I}, which are  usually based on the Sherwood number (Sh) and the Rayleigh number (Ra) as:
\begin{eqnarray}
  \mathrm{Ra}&=& \frac{ \Delta  \rho  k g H}{ \mu  D  \varphi },\\
  \mathrm{Sh}&=& \frac{H}{D  \varphi   \Delta c}F,
\end{eqnarray}
where  \(  \Delta  \rho  \), $k$, $g$, $H$,  \(  \mu  \), $D$,  \(  \varphi  \),  \(  \Delta c \), and $F$ are respectively density difference, permeability, gravitational acceleration, depth, viscosity, diffusion coefficient, porosity, concentration gradient, and dissolution flux. \\

%Knowledge about the \textit{flux growth }regime and \textit{shut-down }regime is restricted to the limited works cited in Table 1 with limitations mentioned above. \
 
 What all these idealized systems have in common is that they consider only a single-phase system where free-phase CO$_2$ as well as multiphase processes that could affect the interface dynamics, partial pressure evolution for each phase, CO$_2$ solubility and the associated density increases  are absent [13, 15, 16, 23, 24].  In addition,  high pressure CO$_2$-brine or CO$_2$-water experiments have been conducted only in blind cells and there is no  visual and quantitative data in the literature on CO$_2$-brine convection dynamics in porous media that can represent the geological carbon sequestration in deep, high-pressure saline aquifers. Here, we present, for the first time, the convective dissolution behavior of CO$_2$ in brine via engineering a  high-pressure Hele-Shaw cell that allows for both qualitative  and quantitative insights into the different flow regimes and the underlying two-phase fluid dynamics. Our visual inspection and scaling analysis of density-driven flow following the dissolution  of free-phase CO$_2$ in an underlying  porous layer  more closely reflect the fate of injected CO$_2$ under  a geologic trap, and will advance our fundamental understanding and predictive capabilities of CO$_2$ sequestration.\\

 Saline aquifers can either be  open, allowing for the compensation of pressure changes by brine migration, or closed (fault bounded), naturally not allowing for   pressure change compensation \citep{akhbari2017causes}.  In closed aquifers, brine saturation with CO$_2$ \citep{hewitt2013convective, slim2013dissolution} and likely pressure drop in  CO$_2$ free phase \citep{wen2018dynamics} would limit the CO\textsubscript{2} dissolution in brine. \citet{wen2018dynamics} showed numerically  that in closed aquifers the dissolution flux in the intermediate, supposedly \textit{quasi-steady }regime is actually not constant, and  suggested the following relation between the flux in closed and open aquifers \citep{wen2018dynamics}:
 \begin{equation}
 F_\mathrm{closed} \approx F_\mathrm{open} C_{s}^{2},
\end{equation}
where  \( C_{s} \)  represents  the equilibrium concentration of  CO$_2$ at the interface with brine. {In this study, confined high-pressure Hele-Shaw cell is designed and implemented in an experimental set-up to investigate the complex behavior of CO$_2$ dissolution in  closed aquifers.}\\

%%%%%%%%%%%%%%%%%%%% Figure/Image No: 1 Ends here %%%%%%%%%%%%%%%%%%%%
 \begin{figure*}
\centerline{\includegraphics[width=\textwidth]{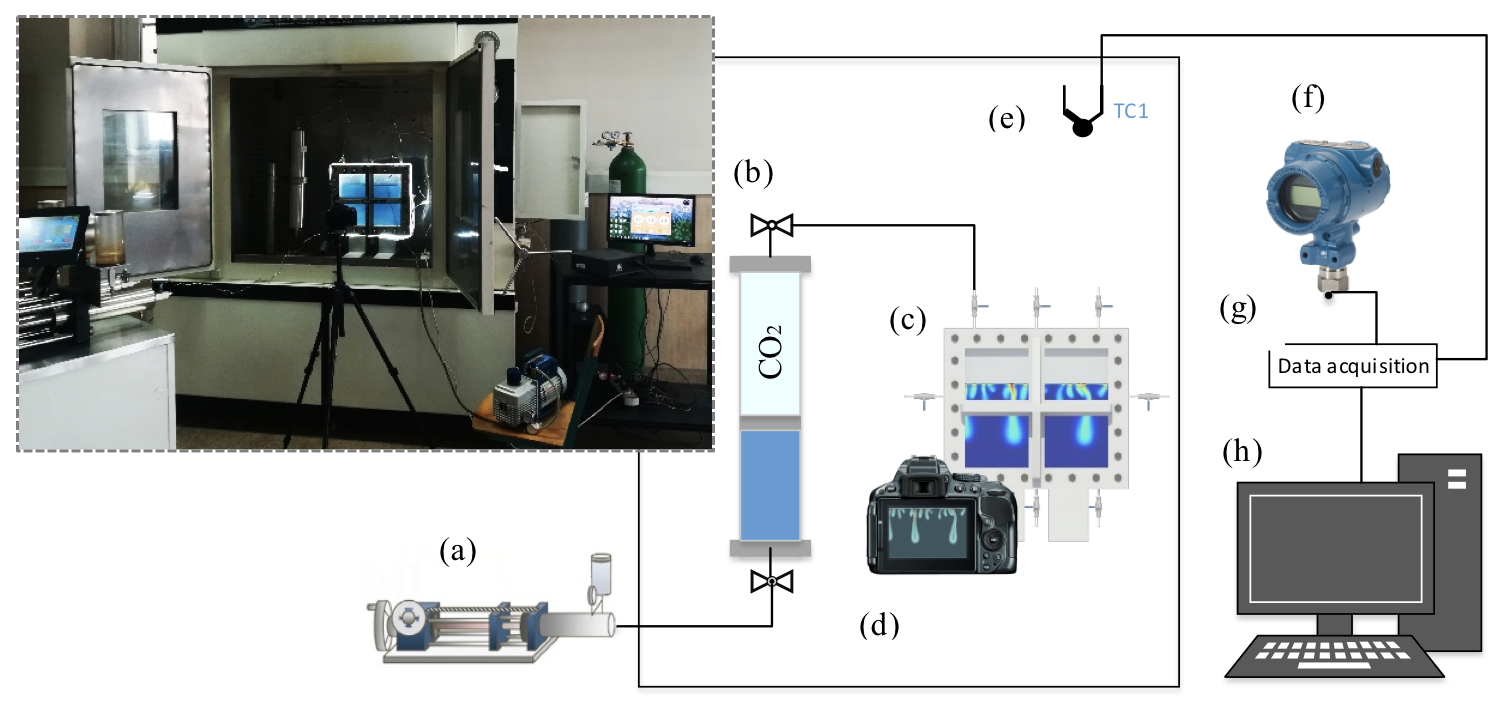}} 
% \makebox[\textwidth][c]{\includegraphics[width=\textwidth]{SI_NB.\ext}}
\caption{{{Experimental set up.  (a) High-pressure pump, (b) PVT cylinder, (c) Hele-Shaw cell, (d) digital camera, (e) thermocouple, (f) pressure gauge, (g) data acquisition system, and (h) a {PC} for data storage.}}}
\label{fig::SI1}
 \end{figure*}
 
{Most previous studies, moreover, have  considered only  \textit{pure} water (i.e., with no salinity) rather than brine. The latter  is typically composed of different salts \citep{aggelopoulos2011interfacial}. Data on brine composition shows the dominance of chloride (Cl\textsuperscript{-}) and sodium (Na\textsuperscript{+}) ions. Among other cations, calcium (Ca\textsuperscript{2+}) is the most frequent one \citep{wellman2003evaluation, gaus2005reactive, xu2006toughreact, bacon2009reactive, azin2013measurement, mohamed2013effect, wang2016three, vu2017changes, shi2018measurement}. } Transition time between the dissolution regimes will control the total amount of dissolved CO\textsubscript{2}  with time. One of our goal here is to provide scaling relations for transition times between dissolution regimes and for dissolution fluxes in CO$_2$ brine systems that contain NaCl and NaCl+CaCl$_2$ mixtures. These relations are helpful for examining system behavior in more realistic conditions where a mixture of salts constitutes the formation brine. \\

The present study is organized as follows. Section II presents the methodology {and  the description of  experimental set-up and details of the implemented tests. }Results are provided and analyzed in section III. {In the first part of this section, the dynamic of dissolution process is examined through the visual data. In the second part, quantitative data (pressure data as well as  dissolution flux) is studied, and scaling relations are obtained for transition times between dissolution regimes and for dissolution flux. }Finally, we draw the conclusions in section IV, followed by Appendix A--E that provide the details for thermodynamic and physical properties of our fluid system.

\section{Methodology} 
We conducted series of eight experiments in high-pressure conditions at different medium permeability and  salinity values. The details of the experiments conducted are given in \textbf{Table II}. {To understand the effect of brine composition we reduced the NaCl amount to 80$\%$  of its initial value, and for the remaining 20$\%$ we added CaCl$_2$. For example, in the corresponding solution of 1 Molal (\textit{M} hereafter) NaCl, we used 0.8 \textit{M} NaCl and 0.1 \textit{M} CaCl$_2$. This way there is similar concentration of {Cl\textsuperscript{-}} in both solutions, and for each divalent cation of {Ca\textsuperscript{2+}} there are two monovalent cations of {Na\textsuperscript{+}} in the corresponding solution. }\\

Experiments have been performed in constant volume of a Hele-Shaw cell that we have designed. The cell is of the internal dimensions of 36$\times$30$\times$2.5 cm (internal volume of 2700 cm$^3$) with aluminum frame covered with a 5 cm thickness plexiglass enabling the visual examination ({the set-up shown in \textbf{Figure1}}). To prevent plexiglass expansion under high-pressure conditions, we attached a steel frame to the main frame. We packed $ \sim $ 23 cm from of total height of 30 cm with glass beads. Permeability range is 400--550 Darcy (D), and porosity is $\sim$ 0.36 for all cases. We used a wider opening (2.5 cm) in our cell, which  allows for adding such \textit{porous} structure into the cell as  an improvement over previous studies. For instance, the onset of convection in bulk fluid experiments happens so instantaneously that usually there is not sufficient data in the early, diffusion-dominated regime. In the presence of porous structure (system with macroscale tortuosity and lower permeability and porosity than bulk fluid), conversely, we can obtain pressure data from diffusion-dominated regime and hence calculate diffusion coefficient from experimental data. Further, our new device has
 negligible boundary effects due to its size, as compared to Hele-Shaw cells used previously. \\

  \begin{figure}
\centerline{\includegraphics[width=\linewidth]{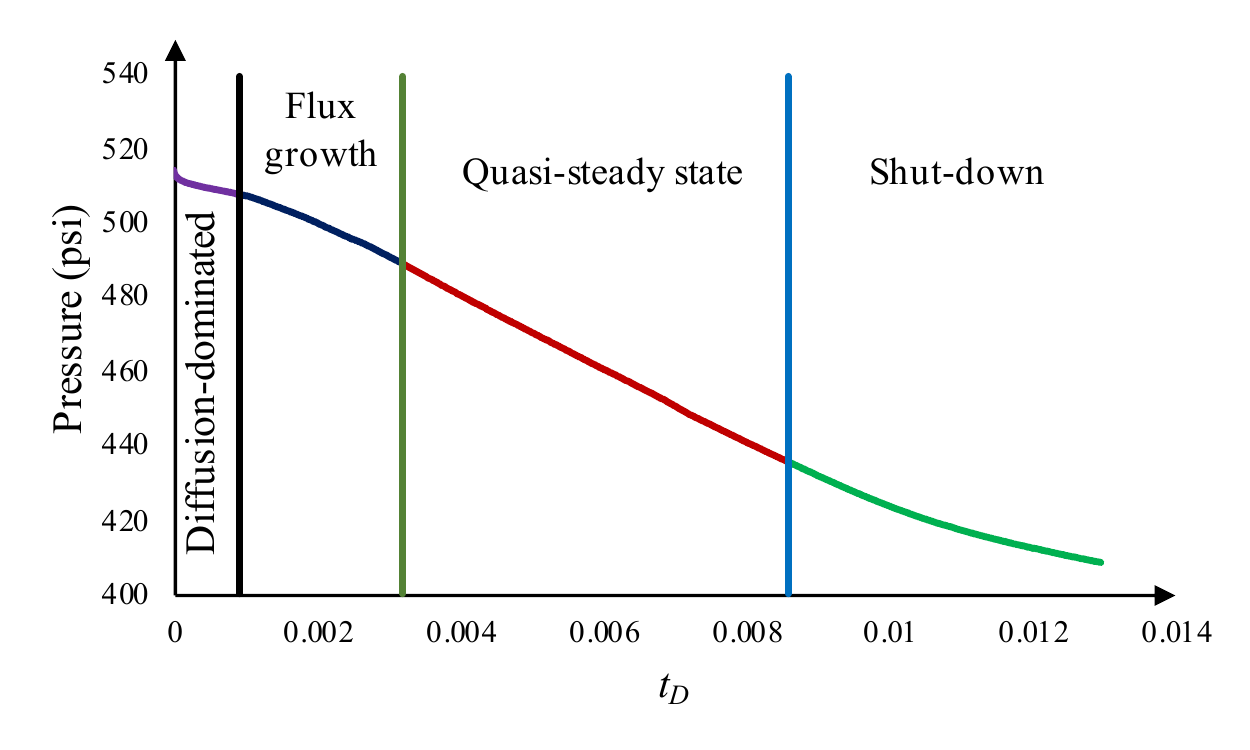}} 
% \makebox[\textwidth][c]{\includegraphics[width=\textwidth]{SI_NB.\ext}}
%\caption{{{Variation of aqueous-phase mass density as a function of pressure and molar fraction of dissolved CO$_2$. Three sample pressures (100, 200, and 300 bar) are shown. Density difference with respect to $\rho_{w,0}$, the pure water density at initial pressure (100 bar), is shown in (a). It is clear that the maximum solubility increases with pressure. The minimum ($\rho_{w,\mathrm{min}}$) and maximum density of aqueous phase ($\rho_{w,\mathrm{max}}$), corresponding respectively to zero and maximum dissolved CO$_2$ composition, are plotted in (b) as a function of pressure; the difference between the two ($\Delta\rho_{w,\mathrm{max}}$), as the main driving force to convection,  is plotted in (c) at each pressure. These results show that the density change due to dissolution is a nonlinear function of the in-situ pressure, and this should be honored.}}}
\caption{Conceptual model for pressure  reduction during dissolution process. Different dissolution regimes are also shown and separated by vertical lines (these boundaries are obtained from dissolution flux analysis)}
\label{fig::SI1}
 \end{figure}

%%%%%%%%%%%%%%%%%%%% Figure/Image No: 2 starts here %%%%%%%%%%%%%%%%%%%%

 \begin{figure*}
\centerline{\includegraphics[width=\textwidth]{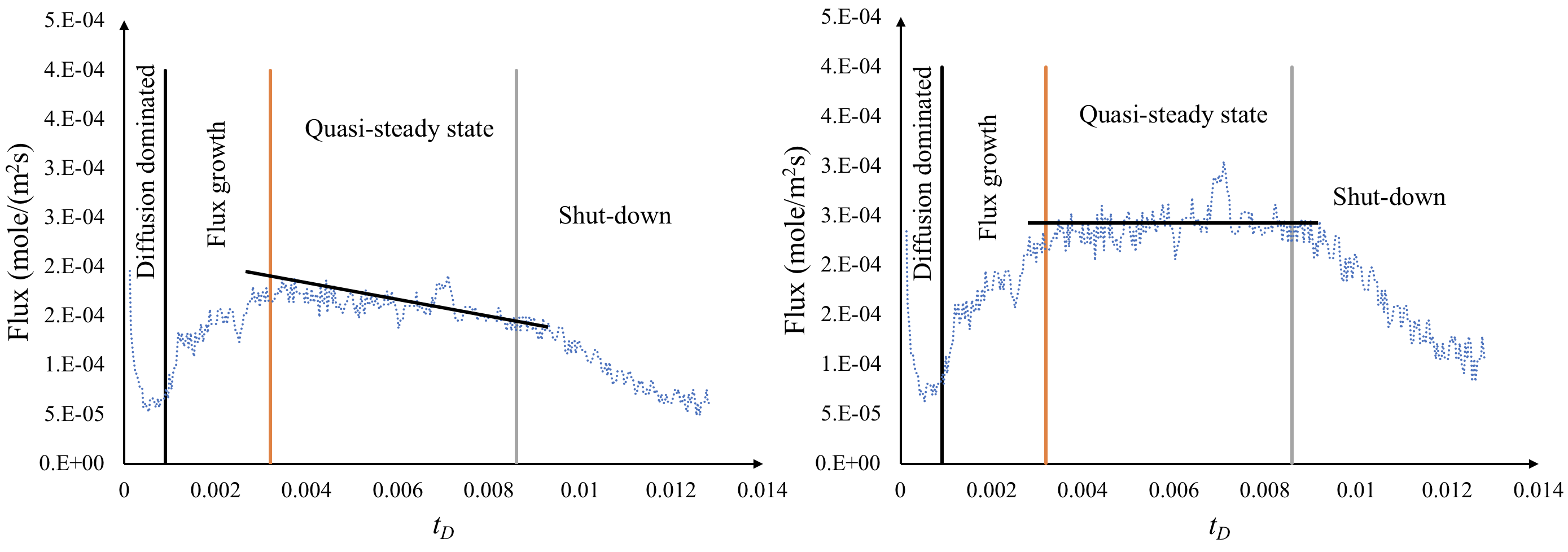}} 
% \makebox[\textwidth][c]{\includegraphics[width=\textwidth]{SI_NB.\ext}}
%\caption{{{Variation of aqueous-phase mass density as a function of pressure and molar fraction of dissolved CO$_2$. Three sample pressures (100, 200, and 300 bar) are shown. Density difference with respect to $\rho_{w,0}$, the pure water density at initial pressure (100 bar), is shown in (a). It is clear that the maximum solubility increases with pressure. The minimum ($\rho_{w,\mathrm{min}}$) and maximum density of aqueous phase ($\rho_{w,\mathrm{max}}$), corresponding respectively to zero and maximum dissolved CO$_2$ composition, are plotted in (b) as a function of pressure; the difference between the two ($\Delta\rho_{w,\mathrm{max}}$), as the main driving force to convection,  is plotted in (c) at each pressure. These results show that the density change due to dissolution is a nonlinear function of the in-situ pressure, and this should be honored.}}}
\caption{Dissolution flux without modification for the closed system (left panel), and  dissolution flux with our new compensation (\(\frac{F_{c}}{C_{s}^{2}}\times C_{s,0}^{2} \approx F \)) for the closed system (right panel).}
\label{fig::SI1}
 \end{figure*}

%%%%%%%%%%%%%%%%%%%% Figure/Image No: 2 Ends here %%%%%%%%%%%%%%%%%%%%

 \begin{figure*}
\centerline{\includegraphics[width=.79\linewidth]{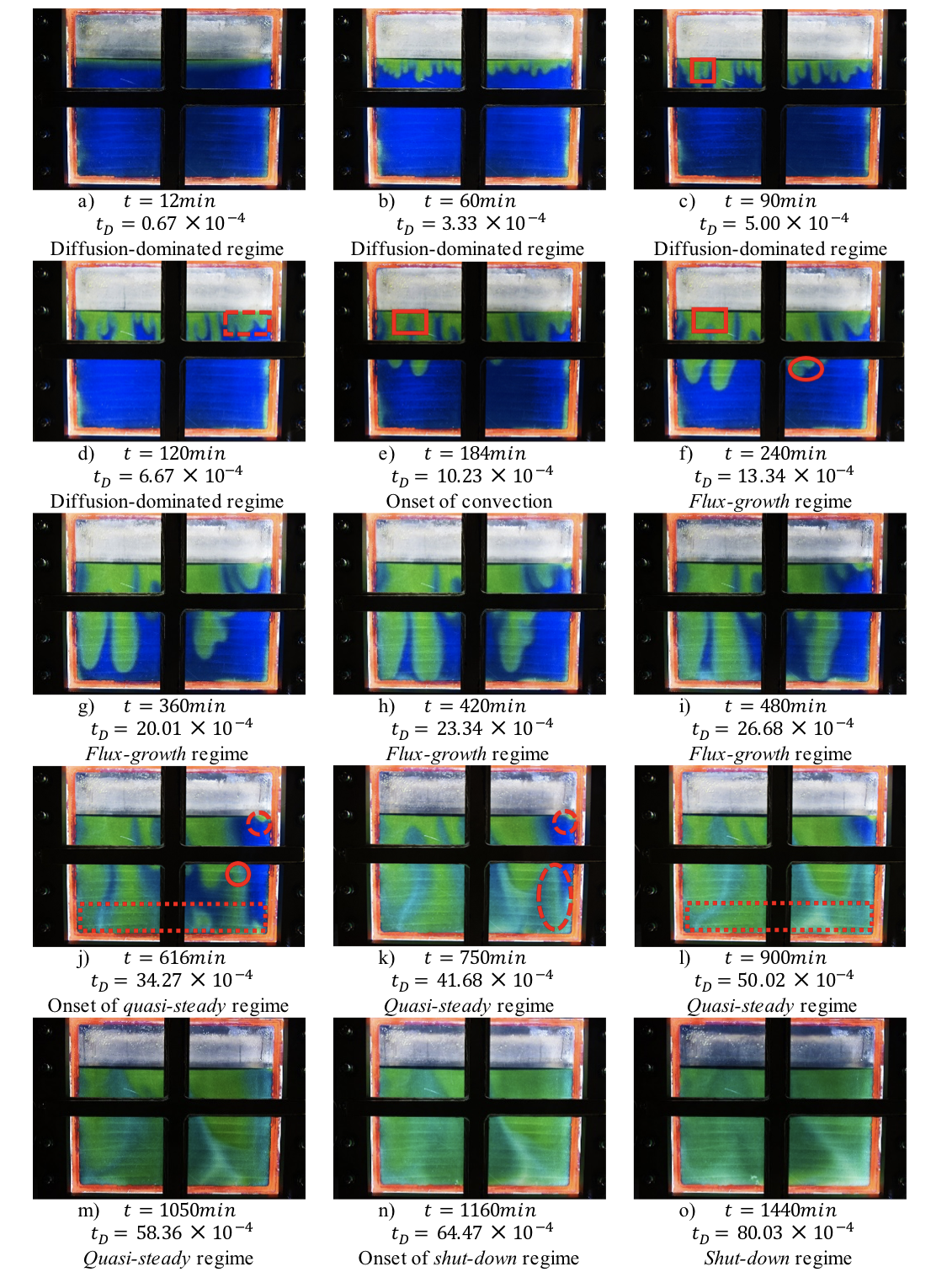}} 
% \makebox[\textwidth][c]{\includegraphics[width=\textwidth]{SI_NB.\ext}}
%\caption{{{Variation of aqueous-phase mass density as a function of pressure and molar fraction of dissolved CO$_2$. Three sample pressures (100, 200, and 300 bar) are shown. Density difference with respect to $\rho_{w,0}$, the pure water density at initial pressure (100 bar), is shown in (a). It is clear that the maximum solubility increases with pressure. The minimum ($\rho_{w,\mathrm{min}}$) and maximum density of aqueous phase ($\rho_{w,\mathrm{max}}$), corresponding respectively to zero and maximum dissolved CO$_2$ composition, are plotted in (b) as a function of pressure; the difference between the two ($\Delta\rho_{w,\mathrm{max}}$), as the main driving force to convection,  is plotted in (c) at each pressure. These results show that the density change due to dissolution is a nonlinear function of the in-situ pressure, and this should be honored.}}}
\caption{The convection patterns for dissolved CO$_2$  in Case 4; the green color shows the dissolved CO$_2$. Symbols are used to show the interactions of convective fingers as follows: square: \textit{side merging}; dashed rectangle: \textit{fading fingers}; rectangle: \textit{root zipping}; oval: \textit{tip splitting}; circle: \textit{necking}; dashed circle: \textit{protoplume reinitiatiion}; dashed oval: \textit{trailing lobe detachment}}
\label{fig::SI1}
 \end{figure*}

%%%%%%%%%%%%%%%%%%%% Figure/Image No: 3 starts here %%%%%%%%%%%%%%%%%%%%

 In our experiments, the porous part of cell is saturated with desired brine solution; we introduce brine into the cell by continuous mixing and gradually adding NaCl or mixture of NaCl and CaCl$_2$ into water. Bromocresol green is used as a pH indicator with 0.02\% weight fraction in brine.  CO$_2$ is injected to the remaining 7 cm in the empty upper section---with no porous structure---to reach the desired pressure. This packing configuration helps to eliminate the capillarity effects as there is no longer a capillary transition zone  between the initial brine- and CO$_2$-saturated  layers. In practice, The cell is placed in an oven and vacuum pump is used to saturate it with brine and the pH indicator solution. Then, the transfer vessel cylinder is loaded with CO$_2$ gas phase. The CO$_2$-containing cell is pressurized, and all equipment in the oven is kept under the 50$^\circ$C for at least 12  hours. Before inserting gas  into the cell, a pump vacuumizes the cell. A valve between gas cylinder and the cell is opened gradually to reduce possible disturbances. Once we obtain a desired pressure, we close the cylinder valve and open the pressure gauge valve.   Initial pressures for the tests are in the range of 502.6--535.3 psi. Because our set-up has constant volume in the isothermal condition ($ \sim $  50 C), the CO$_2$ dissolution into the brine results in some pressure reduction, the continuous monitoring of which can provide a quantitative measure of dissolution. Pressure changes are therefore measured with a high-accuracy pressure gauge (\(\pm  0.1\)  psi resolution). The pressure monitoring and the continuous capturing of  visual snapshots with a digital camera are performed until pressure reaches to an almost constant value. \\

{We use non-dimensional time as  \( t_{D}={tD}/{H^{2}} \)  throughout this study, where $t$, $D$, and $H$ are time, diffusion coefficient and the height of system, respectively. Based on this non-dimensional time, we present an example picture of the dynamic evolution of pressure resulting from our experiments in \textbf{Figure 2 }.}\\

{As mentioned above, we calculate the diffusion coefficient values from the pressure data in the \textit{diffusion-dominated} regime in our experiments, and use these values in this study. By matching the experimental data with simple simulation models following our previous work \citep{mahmoodpour2017design, mahmoodpour2018onset}, we were able to obtain diffusion coefficients. } Unlike in previous experiments with  blind cells where the gas-brine interface is unknown \citep[e.g.,][]{mojtaba2014experimental}, the visual data from our experiments allow us to obtain the volume of CO$_2$\textsubscript{ }gas phase ($V$). We use the equation below to calculate the amount of dissolved CO$_2$ \citep{duan2003improved}:
 \begin{equation}
 n_{\mathrm{diss.}, t}^{\mathrm{CO_2}}= \left( \frac{PV}{zRT} \right) _{i}- \left( \frac{PV}{zRT} \right) _{t},
\end{equation}
where  $n_{\mathrm{diss.}, t}^{\mathrm{CO}_2}$  shows the amount of dissolved CO$_2$ and \textit{i} stands for the initial time condition. The mole fraction of brine in the gas phase is assumed negligible (see \citep{duan2003improved}). \\

The accurate detection of different dissolution regimes is not feasible by pressure data only. Therefore, the dissolution flux $F_c$ is used instead, which is calculated using the values of  dissolved CO$_2$ in brine as follows:
 \begin{equation}
 \frac{\mathrm{d}n_\mathrm{diss.}^\mathrm{CO_2}}{\mathrm{d}t}=AF_{c},
\end{equation}
where ${A}$ is the surface area of the gas-brine interface (here 0.025$\times$0.36 m$^{2}$).\\

To detect the \textit{quasi-steady-similar } regime   in our closed system at the constant volume condition, we follow the analytical approach by \citet{wen2018dynamics}, with a difference that  we  use the  \( C_{s,0}^{2} \)  (initial concentration of CO\textsubscript{2}\ at interface)  in our conversion  to make it non-dimensional as  \(\frac{F_{c}}{C_{s}^{2}}\times C_{s,0}^{2} \approx F\) that can be used for  obtaining  scaling relations. We will discuss this  modification later with details. Since the temperature and salinity are constant, the available correlations for the  solubility of  CO$_2$ in brine can be used to calculate the \textit{C\textsubscript{s}} values at any given pressure. Here, we use the solubility models by \citet{duan2003improved} and \citet{duan2006improved} to calculate the CO$_2$ equilibrium concentration at the gas-brine interface. Importantly, these models show that an increase in salinity will result in a reduction of solubility (i.e., equilibrium concentration). \\
 
 % (simply quasi-steady regime hereafter)

 \textbf{Figure 3} shows the resulting dissolution flux  for the presented pressure curve in \textbf{Figure 2}. It should be noted that, the fluctuating  data of  dissolution flux are converted to produce a smooth curve via  moving average method. Based on the dissolution flux dynamics, the transition between dissolution regimes is detectable. The lines illustrated in the figures show the {onsets of convection}, {quasi-steady (-similar)}, and  shut-down regimes, respectively, from left to right. \\% Images of the process progressing are gathered during tests to examine the process visually. 

While a pressure gauge with a high accuracy of $ \sim $ 0.1 {psi} was used in this study, pressure changes due to CO$_{2}$ dissolution over a small time period on the order of seconds are not detectable. We analyzed  the pressure data based on eight-minute intervals.

\section{Results and Discussion}

Below we present our results for the density-driven flow of CO$_2$ in  brine solutions with dissolved NaCl and NaCl+CaCl$_2$. Specifically, we analyze the dissolution flux and the critical times for different flow regimes. Scaling relations are also presented. To verify the presented scaling relations against  experimental observations, the individual transition times and dissolution rates are further compared with the predicted values.

\subsection{Dynamics of Dissolution}
At early time, the dissolution of CO$_2$ into brine is a \textit{diffusion-dominated} process. The mechanism behind the diffusive regime  is well-documented in the literature; the dissolution flux exhibits the classical Fickian scaling of  $F\propto t^{-0.5} $ \citep{szulczewski2013carbon, soltanian2016critical, soltanian2017dissolution}. We present in  \textbf{Figure 4} the visual results for our Case 4 (introduced in \textbf{Table II}), where the brine has 1 \textit{M }of NaCl  and the permeability is 400 D.  \textbf{Figure 4a} shows the first diffusive regime, where the diffusive layer can be seen as an almost piston-like descending layer near the gas-brine interface. \\

The more dense CO$_2$-rich brine overlying the lighter fresh brine leads to a gravitationally unstable density stratification. Diffusion has damping effect on these instabilities.  As per \citet{szulczewski2013carbon}, in conditions with Ra greater than some critical values (e.g.,  \( Ra>55 \)) instabilities grow with time and result in the development of convective fingers (see \textbf{Figure 4b}). At this time, although the convective fingers are clear, their strength is not sufficient to make substantial changes to the pressure. In early times, convective fingers grow independently. As time passes and convective fingers grow enough, there will be strong interactions between fingers. \\
 
 \begin{figure*}
\centerline{\includegraphics[width=.8\textwidth]{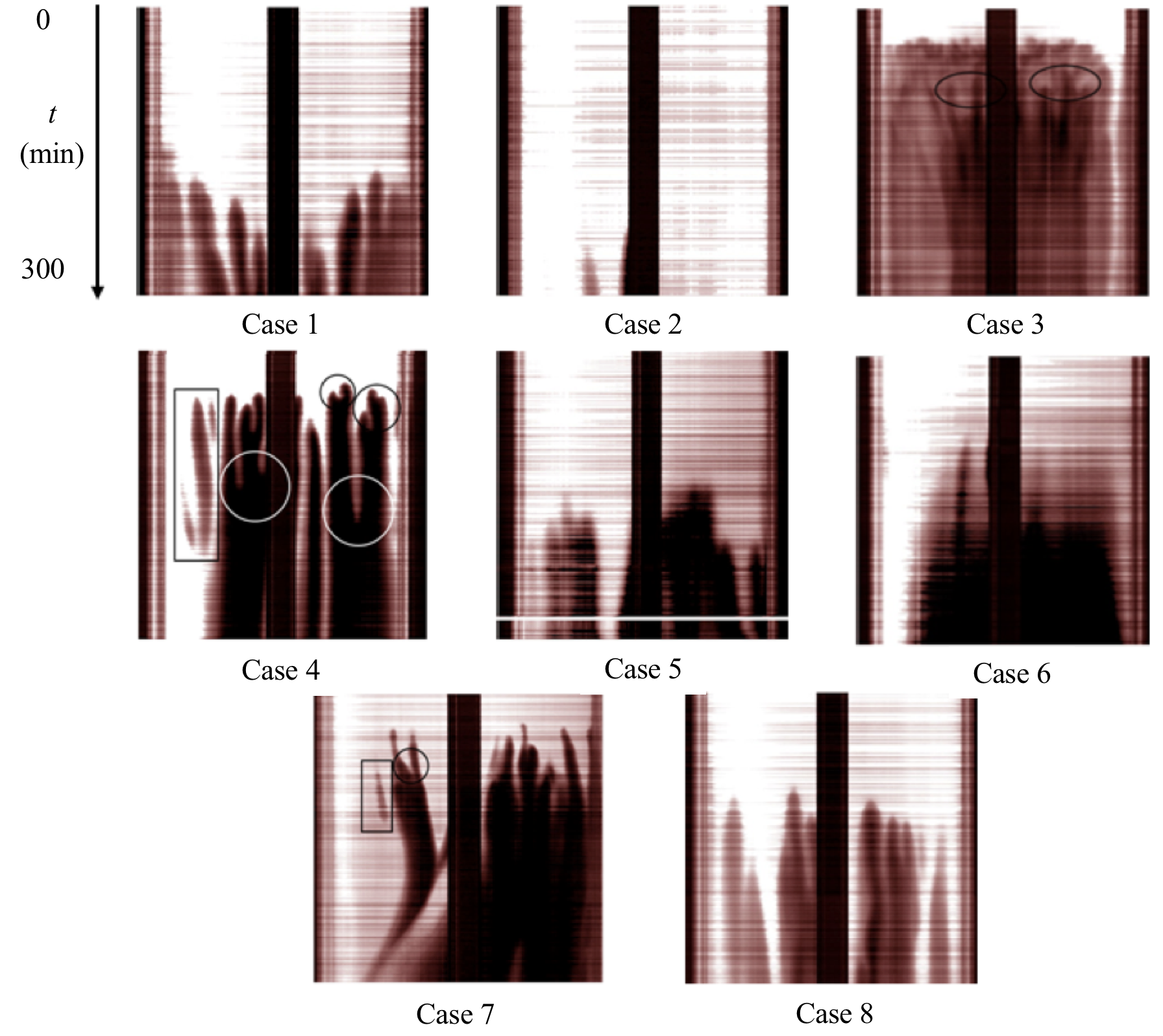}} 
% \makebox[\textwidth][c]{\includegraphics[width=\textwidth]{SI_NB.\ext}}
%\caption{{{Variation of aqueous-phase mass density as a function of pressure and molar fraction of dissolved CO$_2$. Three sample pressures (100, 200, and 300 bar) are shown. Density difference with respect to $\rho_{w,0}$, the pure water density at initial pressure (100 bar), is shown in (a). It is clear that the maximum solubility increases with pressure. The minimum ($\rho_{w,\mathrm{min}}$) and maximum density of aqueous phase ($\rho_{w,\mathrm{max}}$), corresponding respectively to zero and maximum dissolved CO$_2$ composition, are plotted in (b) as a function of pressure; the difference between the two ($\Delta\rho_{w,\mathrm{max}}$), as the main driving force to convection,  is plotted in (c) at each pressure. These results show that the density change due to dissolution is a nonlinear function of the in-situ pressure, and this should be honored.}}}
\caption{Representative spatiotemporal diagrams of the dense fingers (shown by the darker intensity)  at the depth of 5 cm below the gas-brine interface ($x$ axis is the width of cell at that depth while  the $y$ axis is time). Symbols are used to show the behavior of convective fingers  as follows: circle:  the merging of  convective fingers to make a stronger finger; rectangle:   fingers which depart from the dominant finger and diffuse to the nearby stream; oval: shows a condition where the brine rich in dissolved CO$_2$  penetrates  the highly conductive conduit created by previous fingers. {Note that the steel frame used to enhance the capability of  cell, particularly  to tolerate the pressure, is seen with a thick black line in the middle of  pictures}}
\label{fig::SI1}
 \end{figure*}
 
The  mechanisms that control the interactions between gravitational fingers following the dissolution of CO$_2$ in our brine-saturated porous medium are observed as follow. 1) \textit{Side merging} depicted by a square box in \textbf{Figure 4c}: the downward motion of a dominant finger as well as the upward migration of fresh brine from bottom create strong circulating velocity field toward the center of  finger, which in turn drives smaller fingers toward the dominant finger. Eventually, smaller fingers merge into the dominant finger, thus increasing  its strength. 2) \textit{Root zipping} depicted by (solid) rectangular box in \textbf{Figure 4e} and \textbf{Figure 4f}: two fingers  merge from their roots. 3) \textit{Tip splitting} depicted by an oval in \textbf{Figure 4f}:  the tip of a convective finger is flattened by the upward motion of brine and is likely to split into different portions. 4) \textit{Necking }depicted by a circle in \textbf{Figure 4j}: in some part of a finger the width decreases and its feeding from the upper part decreases. 5) \textit{Trailing lobe detachment} depicted by a dashed oval in \textbf{Figure 4k}: a portion of the convective finger is separated from the root. 6) \textit{Reinitiation }depicted by dashed circles in \textbf{Figure 4j} and \textbf{Figure 4k}: new small-scale  fingers, known as protoplumes, are initiated between the dominant descending fingers in intermediate times during  dissolution   with the generation of concentration
gradients below the interface due to upwelling flow of
fresh water. 7) \textit{Fading fingers }which partly is clear\textit{ }in\textit{ }the dashed rectangle in \textbf{Figure 4d}: these fingers are created in early times and grow in a  similar way as the surviving fingers. At some point, the growth of fading fingers  cease,  either when they intersect dominant plumes or through diffusive smearing, and they will act as sourcing pool for feeding  the dominant fingers. Remarkable similar interactions have been reported in prior modeling work on viscous fingering and diffusive-convective dissolution \citep{tan1988simulation, zimmerman1991nonlinear, zimmerman1992viscous, ghesmat2008viscous, ranganathan2012numerical, amooie2017mixing, amooie2017hydrothermodynamic, sabet2018dynamics, amooie2018solutal}.\\

{Under the above mechanisms the dominant convective fingers grow larger and stronger to reach a point at which the pressure decay data show a rapid decline (in comparison to the {diffusion-dominated} regime). }This point is selected as the {onset of convection},  beyond which  the pressure decreases at a greater rate in comparison to the diffusive regime. This time is an important factor from the operational stand-point since there is pressure decline below the caprock. which implies a smaller magnitude of stress and thus a smaller risk of induced leakage. \\

After the {onset of convection} there are two main mechanisms that enhance the dissolution flux. First, as the fingers of CO$_2$-rich brine sink down, fresh brine occupies their place due to mass conservation. Therefore, the concentration gradient at gas-brine interface becomes steeper and mass transfer increases. Second, when fingers sink down, the surface area between sinking fingers and rising fresh brine increases while plume stretching simultaneously steepens
the concentration gradients in the direction perpendicular
to the finger, the combined effect of which leads to  mass transfer enhancement and dissolution across the formation \citep{borgne2014impact}. This forms the flux-growth regime that follows  the {onset of convection}, where the dominant fingers (or megaplumes)  start to appear and will further develop, thereby enhancing the  dissolution flux  (see \textbf{Figures 4f, 4g, 4h, and 4i}). \\%This is associated with a regime known as the {flux growth regime}. 

 %%%%%%%%%%%%%%%%%%%% Table No: 5 starts here %%%%%%%%%%%%%%%%%%%%
 \begin{figure*}
\centerline{\includegraphics[width=\textwidth]{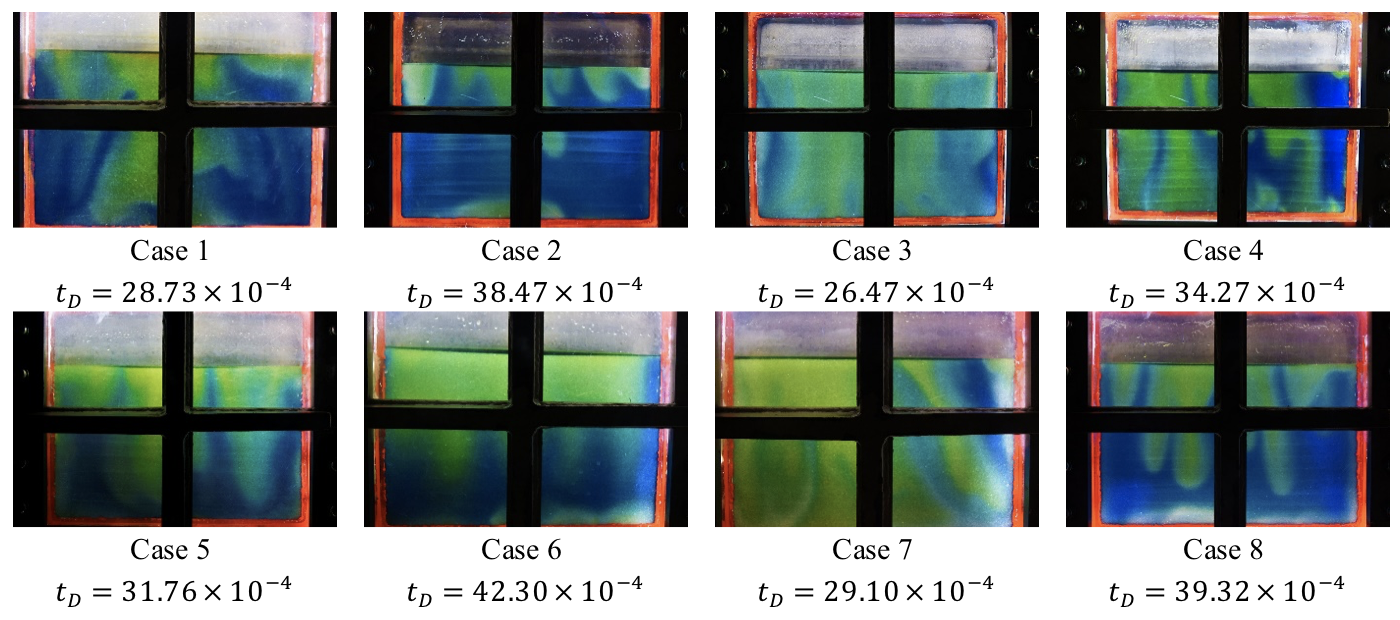}} 
% \makebox[\textwidth][c]{\includegraphics[width=\textwidth]{SI_NB.\ext}}
%\caption{{{Variation of aqueous-phase mass density as a function of pressure and molar fraction of dissolved CO$_2$. Three sample pressures (100, 200, and 300 bar) are shown. Density difference with respect to $\rho_{w,0}$, the pure water density at initial pressure (100 bar), is shown in (a). It is clear that the maximum solubility increases with pressure. The minimum ($\rho_{w,\mathrm{min}}$) and maximum density of aqueous phase ($\rho_{w,\mathrm{max}}$), corresponding respectively to zero and maximum dissolved CO$_2$ composition, are plotted in (b) as a function of pressure; the difference between the two ($\Delta\rho_{w,\mathrm{max}}$), as the main driving force to convection,  is plotted in (c) at each pressure. These results show that the density change due to dissolution is a nonlinear function of the in-situ pressure, and this should be honored.}}}
\caption{Convective fingers at transition times to the {quasi-steady }regime. {Different Ra of these cases results in a different transition time to  the intermediate {quasi-steady }regime}}
\label{fig::SI1}
 \end{figure*}
 
To better elucidate the behavior of the convective fingers resulting in the flux enhancement, a narrow slice is selected at the depth of 5 cm below the gas-brine interface and the formation   of dissolved CO${2}$ patterns over time is captured  by a representative spatiotemporal diagrams of the dense fingers (\textbf{Figure 5}). {Except for Case 6, all cases follow a similar trend: the number of convective fingers and their speed increase with Ra. }These characteristics (fingering number and their speed) are stronger for NaCl solutions in comparison to their corresponding NaCl+CaCl$_2$ mixtures. This is fundamentally because of the smaller diffusion coefficient  for the former (NaCl) than the latter, which would result in larger Ra and thus more pronounced fingering (note that density difference and viscosity profiles for  the corresponding brines of similar salinity are indeed similar to each other). The characteristic features of coarsening dynamics that have been identified in previous numerical studies and experiments on analog fluids are interestingly captured in our CO$_2$-brine experiments at high pressure  as well. As an example,    the Case 3 of  \textbf{Figure 5}  shows that when convective fingers sink down, they create low-resistance pathways where  new high-concentration fluid  preferentially follows the long-lived descending plumes towards the bottom of  system (signified by an oval shape symbol in the figure). As indicated by   circle symbols in the Case 4 of \textbf{Figure 5}, the merging between fingers is more predominant  early on when small-scale  instabilities emerge. Eventually, the continuous merging between neighboring plumes  leads to the development of few larger-scale coherent structures that serve as the conduits for  traveling plumes  (white color circle symbol). The rectangular box  in the Case 4 and 7 represent  those weak convective fingers that fade while separating from a dominant one and diffusing into the main stream.  \\

 Traditionally, when there is no gas cap and pressure change subject to dissolution, the flux grows to a maximum (owing to the mechanisms explained above), beyond which merging and shielding between adjacent elongated fingers combined with diffusive spreading  weaken the concentration gradients in the boundary layer. Hence the dissolution flux stops to grow, and in fact it will decrease transiently until a quasi-steady constant-flux regime develops as a result of a balance between the reinitiation of protoplumes and the coarsening of the existing plumes followed by their subsumption \citep{amooie2018solutal}. Our experiments, however, reveal that such quasi-steady constant-flux regime actually never happens in  a two-phase closed system, as suggested by \textbf{Figure 3}. The pattern coarsening, as the mechanism for flux reduction, dominates the protoplume reinitiation that is the mechanism for flux enhancement (denoted by dashed circles in \textbf{Figures}  \textbf{4j }and\textbf{ 4k}). This is because of the predominant decrease of pressure following the onset of convection, which in turn decreases the solubility of CO$_2$ in brine at the interface and accordingly the dissolution-induced density change, i.e.,  the driving force for  buoyancy-driven convection. The final outcome of this nonlinear dynamics is the decreasing trend, rather than a constant average value, of flux in the intermediate regime before the onset of final shut-down, as predicted theoretically by \citet{wen2018dynamics}.\\
 
 {Despite the decreasing trend of flux in this regime, a  {quasi-steady }regime, however, is attainable in closed systems with the introduced compensation of  dissolution flux as \(\frac{F_{c}}{C_{s}^{2}}\times C_{s,0}^{2} \approx F\)---shown in \textbf{Figure 3}. This is a modification to the scaling suggested by  \citet{wen2018dynamics} as \({F_{c}}/{C_{s}^{2}}\). The results of experiments are analyzed based on the new correction to obtain transition times between dissolution regimes and  the associated scaling relations.} \\

\textbf{Figure 6} shows the  fingering patterns at the  time of transition to the {quasi-steady-similar }regime for all cases. This regime persists until  the upwelling flow starts to carry CO$_2$-rich fluid towards  the gas-brine interface. Next, the concentration gradient at the interface and consequently the dissolution flux decays rapidly  with time, resulting in the shut-down of convective dissolution process. The transition to the convection shut-down regime is presented in \textbf{Figure 4n}, after which the position of  megaplumes is almost constant and the region between adjacent megaplumes  is slowly saturated mainly by diffusion mechanism (\textbf{Figure}  \textbf{4o}). \\

Our findings suggest that the onset of  {shut-down }regime occurs after twice the time required for the first fingers to reach the bottom. The results show that the first contact, e.g., for Case 4 occurs at  almost 480 min after the start of  experiment (\textbf{Figure 4h}), but the {shut-down} regime starts at  approximately  \( 1160>2\times480 \) min. Further analysis of the visual data  explains this behavior. As displayed  in \textbf{Figures}  \textbf{4j} and \textbf{4l} with the dashed rectangular boxes, when the convective fingers reach the \textit{impermeable} bottom of system they initially propagate laterally therein before rising upward with the upwelling flow. The traveling speed of rising wavefront following the impact with bottom boundary is slower than the sinking plumes, partly due to a greater lateral extent (surface area) for the rising fluid  (i.e., a mixture of brine and dissolved CO$_2$) as compared to each sinking finger. Therefore, the onset time of {shut-down} regime is \textit{larger} than the twice the time required for the first impact with  bottom boundary. Note that blind-cell experiments could not observe this. {It should be also noted that the transition times between dissolution regimes are obtained here through quantitative data and will be discussed below with details.}

 \begin{figure}[h]
\centerline{\includegraphics[width=.95\linewidth]{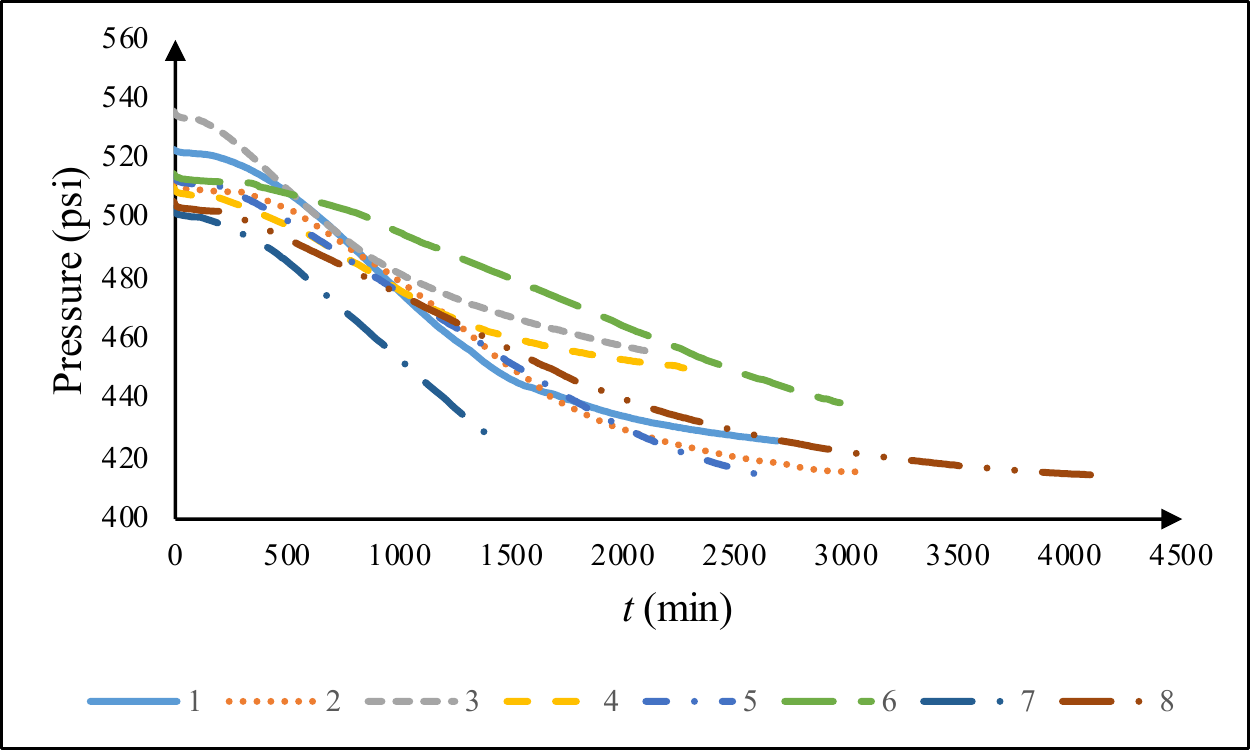}} 
% \makebox[\textwidth][c]{\includegraphics[width=\textwidth]{SI_NB.\ext}}
%\caption{{{Variation of aqueous-phase mass density as a function of pressure and molar fraction of dissolved CO$_2$. Three sample pressures (100, 200, and 300 bar) are shown. Density difference with respect to $\rho_{w,0}$, the pure water density at initial pressure (100 bar), is shown in (a). It is clear that the maximum solubility increases with pressure. The minimum ($\rho_{w,\mathrm{min}}$) and maximum density of aqueous phase ($\rho_{w,\mathrm{max}}$), corresponding respectively to zero and maximum dissolved CO$_2$ composition, are plotted in (b) as a function of pressure; the difference between the two ($\Delta\rho_{w,\mathrm{max}}$), as the main driving force to convection,  is plotted in (c) at each pressure. These results show that the density change due to dissolution is a nonlinear function of the in-situ pressure, and this should be honored.}}}
\caption{Pressure data  obtained from experimental tests}
\label{fig::SI1}
 \end{figure}

\subsection{Scaling Analyses}

{Pressure data  obtained from our tests are presented in \textbf{Figure 7}. During the {diffusion-dominated} regime   pressure (besides temperature) is almost equal for all cases. Therefore,  \textit{salinity} controls the diffusion coefficient and consequently the dissolution rate during the first {diffusion-dominated }regime (see \textbf{Table II}). NaCl+CaCl$_2$-containing brines show a slightly higher diffusion coefficient than the corresponding NaCl brines, most likely  due to the lower total molality for NaCl and CaCl$_2$ mixture as well as the different molecular interactions in different salt solutions (see \textbf{Table II}). In light of this observation, the dissolution rate for NaCl+CaCl$_2$ brine is higher in the first diffusion-dominated regime. The increase in salinity  results in the reduction of  diffusion coefficient, which leads to the lower  dissolution rates in the diffusion-dominated regime, and thus lower pressure decline,  for cases with higher salinity (\textbf{Figure 7}). }\\

{To obtain scaling relations for dissolution flux and the transition times between dissolution regimes, we converted the pressure data to dissolution flux with the help of visual data for the dynamic volume of gas phase. We calculated the modified dissolution flux based on Eq. 3 \citep{wen2018dynamics} for the {quasi-steady-similar }regime to investigate the possible  scaling relations with this character. Note that such  modified  dissolution flux in the {quasi-steady-similar }regime  is independent of time. Results are reported in \textbf{Table III}. }\\

 \begin{figure*}
\centerline{\includegraphics[width=.85\textwidth]{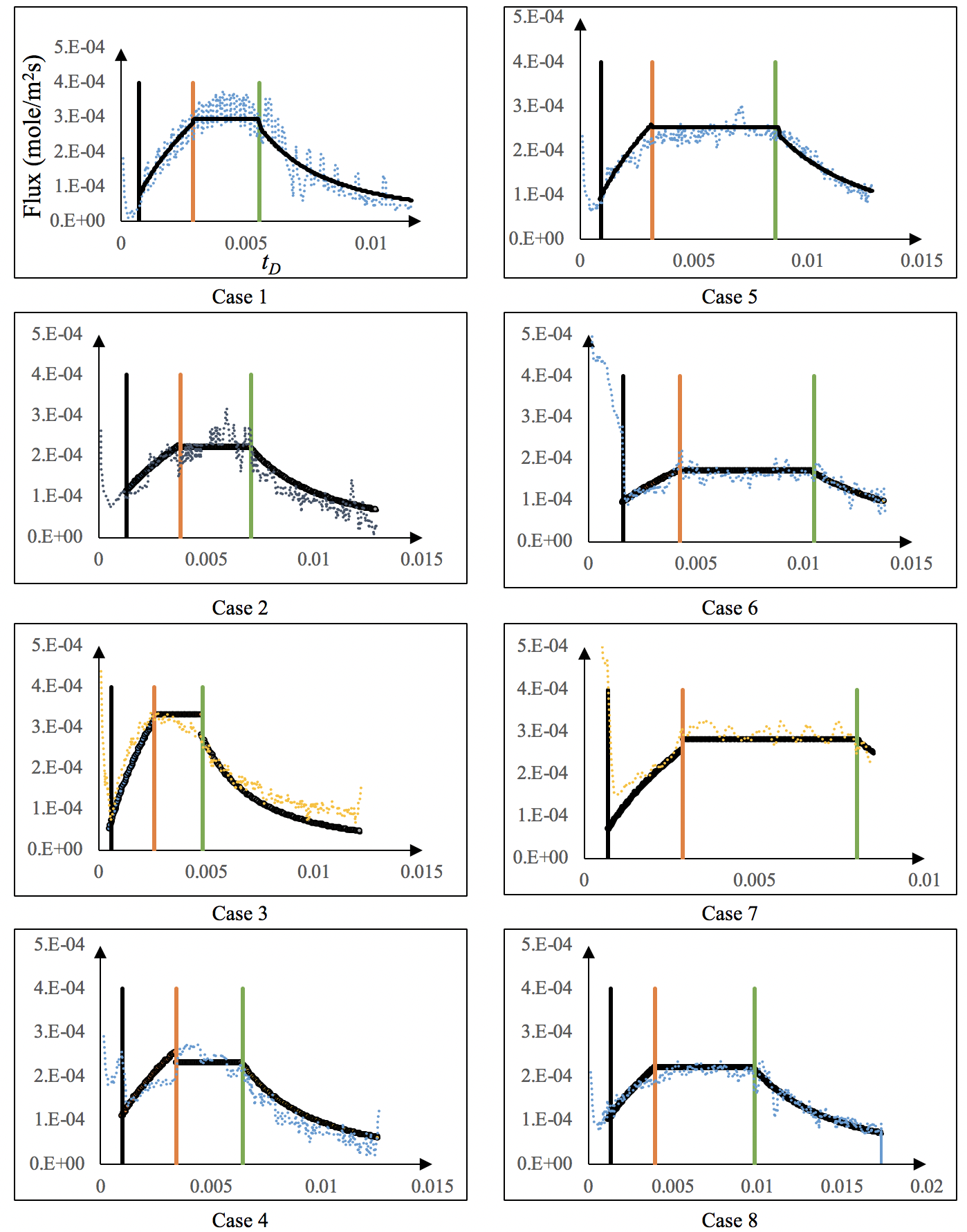}} 
\caption{Compensated dissolution flux  }(\(\frac{F_{c}}{C_{s}^{2}}\times C_{s,0}^{2} \!\approx \! F \) {with dotted line) and fitting curves (continuous line)\  versus dimensionless times for our experiments, which clearly distinguish the quasi-steady regime from early flux-growth and final shut-down regimes.}
\label{fig::SI1}
 \end{figure*}
 
{We find that    \( {F_{c}}/{C_{s}^{2}} \)  surprisingly does not show any specific trend with Ra, unlike the predictions of \citet{wen2018dynamics}. To better understand the reason behind this, let us consider two situations: \textit{i})  Ra is increased through an increase in permeability  (e.g., Case 1 versus  Case 2). In this situation the dissolution flux (${F_c}$) increases with Ra at a given time while  there are no considerable changes in the CO$_2$ solubility values ($C_{s}$); therefore, the overall value of  \({F_{c}}/{C_{s}^{2}} \)  is found to increase with Ra in this situation (e.g., from 5 in Case 2 to 7 in Case 1). \textit{ii}) Ra is increased through a reduction in salinity (e.g., Case 1 versus  Case 3). Whereas the ${F_c}$ increases in this situation, the $C_{s}$  is increased as well (see Section II), and the overall value of  \({F_{c}}/{C_{s}^{2}} \)  actually decreases (from 7 in Case 1 to 4.1 in Case 3). Based on this explanation and opposing situations, we find that \textit{how} Ra changes specifically (via salinity or permeability) is important for the dynamical behavior of  \( {F_{c}}/{C_{s}^{2}} \),   and there is not a general relation between the latter and  Ra.}\\

{Given the definition of Sherwood number as  \(Sh\!=\!\frac{F_cH}{D \varphi  \Delta C/H}\!=\!\frac{F_cH}{D \varphi c_{b,\mathrm{max}}^{s}x_{\mathrm{CO}_2,\mathrm{max}}^{s}/H} \)  \citep{amooie2018solutal}, Sh values  show decreasing trend during the \textit{quasi-steady-similar} regime in a closed system. This is indeed because of the decreasing dynamics of  ${F_c}$ in this regime as shown previously and the other parameters assumed as constant (see \textbf{Figure 3}). Further,   \( c_{b,\mathrm{max}}^{s} \) (constant maximum density of brine)  and  \(x_{\mathrm{CO}_2,\mathrm{max}}^{s} \) (constant maximum solubility fraction)  in the formulation of Sh do change with salinity, and thus  the above-mentioned problem with   \({F_{c}}/{C_{s}^{2}} \) scaling with Ra applies to  Sh scaling as well. }\\

{To overcome these physical challenges in obtaining rigorous scaling relations, we propose a further adjustment   as  \( \frac{F_{c}}{C_{s}^{2}}\times C_{s,0}^{2} \approx F \), where $C_{s,0}^{2}$ is the initial concentration of CO$_2$ at interface and itself sensitive to salinity, such that the dynamic changes in  \( C_{s}^{2} \)  will be compensated with  \( C_{s,0}^{2} \). The results are presented in \textbf{Figure 8} for all Cases. }\\

The analysis of  dissolution flux dynamics shows that for both the solutions of NaCl and NaCl+CaCl$_2$, the dissolution flux increases with time as  \( t_{D}^{0.5} \)  in the {flux-growth} regime. This is consistent with the findings of \citet{newell2018experimental}. Higher Rayleigh numbers result in a greater number of convective fingers with more strength. Therefore, we hypothesize  that the dissolution flux at a given equal time in the flux-growth regime should be greater than that for a lower Ra case, i.e., the pre-factor for the \( t_{D}^{0.5} \) scaling should be positively correlated with Ra. Furthermore, as shown in \textbf{Figure 5} and in supplementary videos \citep{SI}, the number of convective fingers and their strength (quantified with sinking speed) are higher for brine with NaCl than for brine with NaCl+CaCl$_2$ mixtures. Dissolution flux data can quantify these observations. {The relation with the following formula are used to fit the dissolution flux data (with continuous curve):}
 \begin{equation}
 F=m^{F.G} t_{D}^{0.5}-b.
\end{equation}
 
Values of the  \( m^{F.G} \)  and  \( b \)  are reported for each experiment denoted by \( m_{fit.}^{F.G} \)  and  \( b_{fit.}^{F.G} \)  in \textbf{Table III}. {Detailed analysis of these factors prove that they}   interestingly  scale linearly  with Ra  (Eqs. 7 and 8). The estimated values based on these equations are further reported in \textbf{Table III} as  \( m_{est.}^{F.G} \)  and  \( b_{est.}^{F.G} \) .
 \begin{eqnarray}
 m^{F.G}&=& \left( 2.99\times10^{-6} \right) \mathrm{Ra} - 5.48\times10^{-3},\\
  b&=& \left( -7.1\times10^{-8} \right) \mathrm{Ra}+1.83\times10^{-4}.
 \end{eqnarray}

{The modified dissolution flux continues  to  increase until  it plateaus in the quasi-steady period. This time is detected as the onset of {quasi-steady }regime (\(t_{D,fit.}^{Q.S}\)), and is reported in \textbf{Table III}. Results reveal that this onset  occurs earlier for cases with higher Ra in both mixture types. }This is because the interactions between convective fingers are more intense at higher Ra, and therefore   the megaplumes form earlier during the experiments. Based on the same reasoning, NaCl solutions show earlier onset for the {quasi-steady} regime in comparison with their corresponding  NaCl+CaCl$_2$ solutions. {Detailed analysis of the onset of {quasi-steady }regime show that they correlate through the following equations with Ra in two mixtures:}
 \begin{eqnarray}
 t_{D, \mathrm{NaCl}}^{Q.S}&=&5.437\times \mathrm{Ra}^{-0.899},\\
 t_{D, \mathrm{NaCl+CaCl}_2}^{Q.S}&=&13.591\times \mathrm{Ra}^{-1.01}.
 \end{eqnarray}
{Estimated times for onset of {quasi-steady }regime (\(t_{D,est.}^{Q.S}\)) with the proposed scaling relations are presented in \textbf{Table III}, which show good agreements with the fitted values. As discussed in \textbf{Figure 4}, protoplume reinitiation  helps to maintain the  dissolution flux during {quasi-steady }regime. Cases with higher Ra show higher (modified) dissolution flux during this regime. }These constant dissolution rates (which are presented with  \( a_{fit.}^{Q.S} \) ) are fitted with the Eq. 11 for all cases and the resulting estimations are reported with the symbol of  \( a_{est.}^{Q.S} \) .
 \begin{equation}
  F=a^{Q.S}= \left( 2.30 \times 10^{-8} \right) \mathrm{Ra}^{1.126}.
 \end{equation}
 The almost linear scaling of compensated dissolution flux with Ra obtained from our novel experiments is somewhat striking given the debate that  exists for the flux scaling (linear versus sublinear) in the quasi-steady regime (see \textbf{Table I}).\\

 %has a longer duration Because the onset of  {shut-down }regime naturally occurs in  later times than the onset of {quasi-steady }regime; therefore, it is also expected to see a larger separation between   cases of {NaCl }and NaCl+CaCl$_2$ solutions for the onset of  {shut-down}.
Dissolution process continues in the {quasi-steady }regime until the wavefront of CO$_2$-rich brine from the bottom reaches the interface. It is expected to find an earlier onset of  {shut-down }regime for higher Ra cases just because of its control on the traveling speed of convective fingers. We observe that our experimental results meet this  expectation. Another key feature  we find is that NaCl+CaCl$_2$ solutions show longer duration of  {quasi-steady }regime as compared to the corresponding NaCl solutions (\textbf{Figure 8}). This can be justified by our previous explanation as to why the onset of shut-down takes more than two times the time of the first impact with bottom boundary; lower Ra for the NaCl+CaCl$_2$ solutions compared to their corresponding NaCl brine would result in even longer travel times for the rising wavefront. Hence, the onset of shut-down occurs much later for the NaCl+CaCl$_2$ than for an equivalent NaCl solution, accordingly with  longer duration of the quasi-steady regime. The analysis of  the onset of {shut-down }times (\(t_{D,fit.}^{Sh.}\)) shows that following relations describe well the transition times. {The corresponding estimated values (\(t_{D,est.}^{Sh.}\)) with these relations are reported in \textbf{Table III} and show close agreement with the fitted values.}
 \begin{eqnarray}
 t_{D, \mathrm{NaCl}}^{Sh.}&=&12.874\times \mathrm{Ra}^{-0.928},\\
t_{D, \mathrm{NaCl+CaCl}_2}^{Sh.}&=&18.659\times \mathrm{Ra}^{-0.929}.
  \end{eqnarray}

During the {shut-down }regime, the CO$_2$ concentration gradient at gas-brine interface constantly decreases while reducing the dissolution flux. Whereas the previous theoretical and numerical studies reported a  \( F\propto t^{-2} \)  characteristic behavior during this regime \citep{slim2014solutal, de2017dissolution, wen2018dynamics}, recent experiments of \citet{newell2018experimental} suggested a  \( F\propto t^{-1.75} \) scaling  during the {shut-down }regime. Our experiments remarkably confirmed the temporal scaling behavior of  the theoretical and numerical studies, where  \( F=a^{S.D}t_{D}^{-2} \)  describes our results (see the continuous line during this regime in \textbf{Figure 8}). Pre-factors of  \( a^{S.D} \)  {are also reported in \textbf{Table III}. These values, however, do not show a distinct trend with Ra, and it seems  they are a function of  diffusion coefficients and depend on the history of prior regimes. Cases with higher salinity show larger values for pre-factor, because of their lower concentration of dissolved CO$_2$  as a result of prior dissolution events. The NaCl+CaCl$_2$ mixtures show higher pre-factors in comparison to their corresponding  NaCl solutions because of larger diffusion coefficients.}

\section{Discussion and Conclusions}

In this study, high-pressure laboratory Hele-Shaw experiments are performed to examine the    convective dissolution  of actual CO$_2$ in a closed porous media system saturated with  brines that contain  NaCl and NaCl+CaCl$_2$ mixtures at different molality. The common concerns of previous studies that use analog fluid systems to emulate the CO$_2$-brine behavior, such as non-monotonic density profile and miscibility of system, are resolved in our work. Further, in comparison to the experiments performed in  blind cells, visual data are available in this study enabling us to   examine the dynamics of convective fingers during dissolution. Performing the experiments at high pressure not only more closely reflects the subsurface conditions but also makes it possible to obtain diffusion coefficients and dissolution fluxes through the measurable pressure changes during the dissolution process.  Our new engineering   provides an unprecedented opportunity to bridge the gaps among the visual low-pressure and high-pressure blind cell \textit{bulk}-fluid experiments and the analog fluid experiments in  \textit{porous} Hele-Shaw cells at atmospheric conditions. Juxtaposing the qualitative and quantitative data in this study further decreases the possible errors during data analysis.\\

Qualitative data (image) analyses reveal that in the early times, the dissolution process is {diffusion-dominated}. The dissolution of CO$_2$ into underlying brine creates a flat diffusive boundary layer under the gas-brine interface. The mixture of CO$_2$ and brine overlying the lower density fresh brine leads to a  gravitationally unstable stratification. When the diffusive boundary layer is thickened enough, instabilities initiate and grow  as  convective fingers. Following their appearance,  convective fingers grow independently, early on, but as time goes on they interact more with each other and create  stronger fingers. Seven mechanisms for  fingering interactions are identified in this study as follows: \textit{side merging; root zipping; tip splitting; necking; trailing lobe detachment; protoplume reinitiation; fading fingers}. We found that the number of convective fingers, their speed and interactions are higher in NaCl solutions as compared to the corresponding NaCl+CaCl$_2$ solutions. In each type of solutions,  the speed of fingers and their interactions are positively correlated with Ra. Increasing salinity results in a  decrease of  convective flow strength. After some period of time, convective fingers well develop into larger-scale coherent structures and {quasi-steady-similar }regime starts where  new smaller scale fingers emerge  at the gas-brine interface helping the maintenance of   dissolution flux. However, the considerable reduction in gas pressure following CO$_2$ dissolution, especially after the onset  of convection, results in a reduction in solubility and concentration at the gas-brine interface. The flux-enhancing protoplume reinitiation is thus reduced, such that it is no longer sufficient to balance the flux-reducing coarsening mechanisms. Final outcome is a progressively decreasing trend in flux within the intermediate regime before the onset of shut-down regime. The latter begins when the  brine brought to the  interface starts to contain dissolved CO$_2$ in late times, following which the position of  megaplumes is almost constant and the region in between  is slowly saturated mainly through diffusion. The onset of {shut-down }regime occurs earlier for {NaCl }solutions.\\

Our experiments are conducted in constant volume condition. Consequently, the dissolution of CO$_2$ into brine is accompanied by a reduction of  the pressure of  gas phase. The analysis of pressure data  provides a unique quantitative measure of the  dynamic process. To detect the transition between dissolution regimes and reveal the scaling behavior of dissolution, the dissolution fluxes are calculated from the pressure curves with the help of visual data for gas volume. {Further, we  introduced a modification to dissolution flux in order to yield insights into the scaling of  {quasi-steady }regime (similar to open systems) as a function of  Ra. }With the conversion of  \(\frac{F_{c}}{C_{s}^{2}}\times C_{s,0}^{2} \approx F \), dissolution fluxes in the closed system appear to become constant  during the {quasi-steady }regime. Using this definition of flux,  four dissolution regimes are identified as follows: {diffusion-dominated}; {flux-growth}; {quasi-steady} and {shut-down }regimes. Scaling relations based on the dimensionless numbers are introduced for the transition times between these regimes. All transition times for the quasi-steady and shut-down regimes for different salt types exhibit an inverse proportionality with Ra. However, the comparison between results reveals that NaCl solutions show earlier onsets of {quasi-steady} and {shut-down }regimes as compared with those from the  corresponding NaCl+CaCl$_2$ solutions. The difference  between the onset times of {shut-down} in the two types of brine  is larger than those of the {quasi-steady }regime. Therefore, NaCl+CaCl$_2$ systems experience a longer period in the {quasi-steady }regime. Dissolution flux in all cases follows the    \(F\propto t_{D}^{0.5}\)  and  \( F\propto t_{D}^{-2} \) scaling behavior for the {flux-growth} and {shut-down }regimes, respectively. During the {flux-growth {regime}}, the pre-factors for \( F\propto t_{D}^{0.5} \) for NaCl solutions are higher than those for NaCl+CaCl\textsubscript{2} solutions, and  both linearly scale with Ra.  Further, {NaCl }solutions show  higher dissolution fluxes than NaCl+CaCl$_2$ mixtures during the {quasi-steady }regime, with the \textit{compensated} dissolution fluxes almost linearly scaling with Ra irrespective of salt types.  For the {shut-down} regime, the dissolution rates for NaCl+CaCl$_2$ mixtures are also higher.\\% These pre-factors for flux-growth and dissolution flux for the {quasi-steady }regimes are correlated with Ra. \\

Findings of this study advance our knowledge of  CO$_2$  dissolution into formation brine following  geological sequestration in deep saline aquifers. Scaling relations are introduced based on the dimensionless numbers, which enable us to use them in similar studies and possibly field experiments. Since  CaCl$_2$ is the most common salt after  NaCl in the majority of aquifers, this work helps to obtain a more realistic estimation for the short- and long-term fate and transport of CO$_2$---particularly in terms of the dissolution flux and transition times between critical regimes---following its storage in the subsurface. %Also, most of the previous studies had been done based on the NaCl solution; therefore, introduced scaling relations can be used to convert them to more realistic cases.

\appendix
\section{Solubility of CO$_2$ in brine}

Solubility of CO$_2$ in brine can be obtained from \citet{duan2003improved, duan2006improved}:
\begin{widetext}
\begin{align}
\mathrm{ln}x_{\mathrm{CO}_2}=\mathrm{ln} \left( y_{\mathrm{CO}_2} \phi _{\mathrm{CO}_2}P \right) -\frac{ \mu _{\mathrm{CO}_2}^{1 \left( 0 \right) }}{RT}-2\lambda _{\mathrm{CO}_2-\mathrm{Na}} \left(m_{\mathrm{Na}}+m_{\mathrm{K}}+2m_{\mathrm{Ca}}+2m_{\mathrm{Mg}} \right) - \notag \\- \zeta _{\mathrm{CO}_2-\mathrm{Na}-\mathrm{Cl}}m_{\mathrm{Cl}} \left( m_{\mathrm{Na}}+m_{\mathrm{K}}+m_{\mathrm{Mg}}+m_{\mathrm{Ca}} \right) +0.07m_{\mathrm{SO4}},
\end{align}
where \textit{T} in \textit{K} and pressure in bar.  \(  \mu _{{\mathrm{CO}_2}}^{1 \left( 0 \right) } \) ,  \(  \lambda _{{\mathrm{CO}_2}-\mathrm{Na}} \)  and  \(  \zeta _{{\mathrm{CO}_2}-\mathrm{Na}-\mathrm{Cl}}m_{\mathrm{Cl}} \)  are standard chemical potential, interaction parameter between CO$_2$ and {Na\textsuperscript{+}} and interaction parameter between CO$_2$, {Na\textsuperscript{+}} and {Cl\textsuperscript{-}} respectively. Following equation is used to calculate these parameters:
 \begin{equation}
 \mathrm{Par} \left( T,P \right) =c_{1}+c_{2}T+\frac{c_{3}}{T}+c_{4}T^{2}+\frac{c_{5}}{ \left( 630-T \right) }+c_{6}P+c_{7}P\mathrm{ln}T+c_{8}\frac{P}{T}+c_{9}\frac{P}{ \left( 630-T \right) }+c_{10}\frac{P^{2}}{ \left( 630-T \right) ^{2}}+c_{11}T\mathrm{ln}P
\end{equation}

The fugacity coefficient ($\phi _{{\mathrm{CO}_2}}$)  is calculated from the following non-iterative equation:
  \begin{equation}
 \phi _{{\mathrm{CO}_2}}=c_{1}+ \left[ c_{2}+c_{3}T+\frac{c_{4}}{T}+\frac{c_{5}}{T-150} \right] P+ \left[ c_{6}+c_{7}T+\frac{c_{8}}{T} \right] P^{2}+\frac{ \left[ c_{9}+c_{10}T+\frac{c_{11}}{T} \right] }{\mathrm{ln}P }
 \end{equation}

Constants of \textit{c\textsubscript{1}- c\textsubscript{15}} are presented in \textbf{Table IV}.\par
\end{widetext}

\section{Viscosity of NaCl solution \citep{mao2009viscosity}}

\begin{equation}
 \mathrm{ln} \left( \frac{ \mu _\mathrm{brine~of~CO_2}}{ \mu _\mathrm{H_2O}} \right) =Am+Bm^{2}+Cm^{3},
\end{equation}
where \textit{A}, \textit{B} and \textit{C} are functions of temperature (in \textit{K}):\par
\begin{eqnarray}
 A&=& c_{1}+c_{2}T+c_{3}T^{2}, \\
 B&=&c_{4}+c_{5}T+c_{6}T^{2},\\ 
 C&=&c_{7}+c_{8}T. 
\end{eqnarray}

\textit{c\textsubscript{1}- c\textsubscript{8}} are obtained from the experimental data and available in \textbf{Table V}.\par

\section{Viscosity of {CaCl\textsubscript{2}+ NaCl} solution}
To obtain the viscosity of the {NaCl+ CaCl\textsubscript{2}} solution, the simple mixing rule is used \citep{zhang1997viscosity}:
\begin{widetext}
\begin{eqnarray}
  \Delta  \mu _\mathrm{mix.}&=& \Delta  \mu _\mathrm{CaCl_2}+ \Delta  \mu _\mathrm{NaCl}, \\
   \mu _{\mathrm{mix.}}&=& \mu _\mathrm{H_2O} \left( 1+ \Delta  \mu _\mathrm{mix.} \right),\\
  \Delta  \mu _\mathrm{NaCl}&=&c_{1}m_\mathrm{NaCl}^{0.5}+c_{2}m_\mathrm{NaCl}+c_{3}m_\mathrm{NaCl}^{2}+c_{4}m_\mathrm{NaCl}^{3.5}+c_{5}m_\mathrm{NaCl}^{7},\\
  \Delta  \mu _\mathrm{CaCl_2}&=&c_{6}m_\mathrm{CaCl_2}^{0.5}+c_{7}m_\mathrm{CaCl_2}+c_{8}m_\mathrm{CaCl_2}^{2}+c_{9}m_\mathrm{CaCl_2}^{3.5}+c_{10}m_\mathrm{CaCl_2}^{7},
\end{eqnarray}
\end{widetext}
where $c_1$--$c_{10}$ are calculated from the experimental data on the specific concentration of the salts. In the required concentrations, parameters are interpolated. For 1 \textit{M}. concentration of brine, these values are presented in \textbf{Table V}.

\section{Brine density}

NaCl solution density \citep{potter1977volumetric}:
\begin{equation}
   \rho _{b}=\frac{1000 \rho _{w}+M_\mathrm{NaCl}m_\mathrm{NaCl} \rho _{w}}{1000+A_{0}m_\mathrm{NaCl} \rho _{w}+B_{0}m_\mathrm{NaCl}^{1.5} \rho _{w}+C_{0}m_\mathrm{NaCl}^{2} \rho _{w}}
\end{equation}

To obtain the {NaCl+ CaCl\textsubscript{2}} solution density, the relative of the {NaCl+ CaCl\textsubscript{2}} mixture to the NaCl solution is calculated by Al Ghafri et al. model \citep{al2012densities}:
\begin{widetext}
\begin{eqnarray}
  \rho _{b} \left( T,P,m \right) &=&  \rho_\mathrm{ref} \left( T,m \right)   \left[ 1-C(m) \mathrm{ln}\left (\frac{B \left( T,m \right) +P}{B \left( T,m \right) +P_\mathrm{ref} \left( T \right) } \right)  \right] ^{-1},\\
  \mathrm{ln}\frac{P_\mathrm{ref} \left( T \right) }{P_{c}}&=& \left( \frac{T_{c}}{T} \right)  \{  \sigma _{1} \varphi + \sigma _{2} \varphi ^{1.5}+ \sigma _{3} \varphi ^{3}+ \sigma _{4} \varphi ^{3.5}+ \sigma _{5} \varphi ^{4}+ \sigma _{6} \varphi ^{7.5} \},
\end{eqnarray}
where \textit{T\textsubscript{c}} and \textit{P\textsubscript{c}} are critical values and  $\varphi = 1-{T}/{T_{c}}$, and 
\begin{eqnarray}
   \rho_\mathrm{ref} \left( T,m \right) - \rho _{0} \left( T \right) &=& \sum _{i=1}^{i=3} \alpha _{i0}m^{ \left( i+1 \right) /2}+ \sum _{i=1}^{i=3} \sum _{j=1}^{j=3} \alpha _{ij}m^{ \left( i+1 \right) /2} \left( \frac{T}{T_{c}} \right) ^{ \left( j+1 \right) /2},\\
   B \left( T,m \right) &=& \sum _{i=0}^{i=1} \sum _{j=0}^{j=3} \beta _{ij}m^{i} \left( \frac{T}{T_{c}} \right) ^{j},\\
 C \left( m \right) &=& \gamma _{0}+ \gamma _{1}m+ \gamma _{2}m^{1.5},\\
   \frac{ \rho _{0} \left( T \right) }{ \rho _{c}}&=&1+c_{1} \varphi ^{\frac{1}{3}}+c_{2} \varphi ^{\frac{2}{3}}+c_{3} \varphi ^{\frac{5}{3}}+c_{4} \varphi ^{\frac{16}{3}}+c_{5} \varphi ^{\frac{43}{3}}+c_{6} \varphi ^{\frac{110}{3}},
\end{eqnarray}
where ${c_{1}}=1.992741, {c_{2}}=1.099653, {c_{3}}=-0.510839, {c_{4}}=-1.754935, {c_{5}}=-45.517035, {c_{6}}=674694.45, {\beta_{00}}=-1622.40, {\beta_{01}}=9383.80, {\beta_{02}}=-14893.80, {\beta_{03}}=7309.10, {\gamma_{0}}=0.11725, {\sigma_{1}}=-7.859518, {\sigma_{2}}=1.844083, {\sigma_{3}}=-11.786650, {\sigma_{4}}=22.680741, {\sigma_{5}}=-15.961872,$ and $\sigma_{6}=1.801225$. Required parameters for NaCl and CaCl$_2$ are presented in \textbf{Table VI}.
 \end{widetext}

\section{Partial molar volume of CO$_2$ in water \citep{garcia2001density}}
\begin{equation}
  V_\mathrm{CO_2}=37.51-9.585\times10^{-2}T+8.740\times10^{-4}T^{2}-5.044\times10^{-7}T^{3},
\end{equation}
where ${T}$ is in degrees Celsius and  $V_\mathrm{CO_2}$  is in cm\textsuperscript{3}/mole.
%{\fontsize{0pt}{0.0pt}\selectfont \uline{ }\par}\par

%\bibliography{bibliography}
% \bibliography{bibfile}

%\printbibliography

\pagebreak
{
\setlength\extrarowheight{-4pt}
\begin{table*}[h]
 \centering
 \caption{A summary of scaling relations for estimating  dissolution flux. \\\textsuperscript{$\ast$ }{\fontsize{10pt}{12.0pt}\selectfont Methanol and ethylene-glycol\ \ \ \  \textsuperscript{$\ast$$\ast$ }Propylene glycol\ \ \ \  \textsuperscript{$\ast$$\ast$$\ast$}a, a$'$ are constant values}}
\begin{tabular}{p{0.82in}p{0.59in}p{0.65in}p{0.81in}p{0.75in}p{1.74in}}
\hline
 
%row no:1
\multicolumn{1}{|p{0.82in}}{\Centering {\fontsize{10pt}{12.0pt} \selectfont Reference}} & 
\multicolumn{1}{|p{0.59in}}{\Centering {\fontsize{10pt}{12.0pt}\selectfont System}} & 
\multicolumn{1}{|p{0.65in}}{\Centering {\fontsize{10pt}{12.0pt}\selectfont Regime}} & 
\multicolumn{1}{|p{0.81in}}{\Centering {\fontsize{10pt}{12.0pt}\selectfont Analysis method}} & 
\multicolumn{1}{|p{0.75in}}{\Centering {\fontsize{10pt}{12.0pt}\selectfont Ra range}} & 
\multicolumn{1}{|p{1.74in}|}{\Centering {\fontsize{10pt}{12.0pt}\selectfont Equation}} \\
\hhline{------}
%row no:2
\multicolumn{1}{|p{0.82in}}{\cellcolor[HTML]{FFFFFF}\Centering {\fontsize{10pt}{12.0pt}\selectfont \citep{hesse2008mathematical}}} & 
\multicolumn{1}{|p{0.59in}}{\cellcolor[HTML]{FFFFFF}\Centering {\fontsize{10pt}{12.0pt}\selectfont -}} & 
\multicolumn{1}{|p{0.65in}}{\cellcolor[HTML]{FFFFFF}\Centering {\fontsize{10pt}{12.0pt}\selectfont \textit{Quasi-steady}}} & 
\multicolumn{1}{|p{0.81in}}{\cellcolor[HTML]{FFFFFF}\Centering {\fontsize{10pt}{12.0pt}\selectfont Numerical}} & 
\multicolumn{1}{|p{0.75in}}{\cellcolor[HTML]{FFFFFF}\Centering {\fontsize{10pt}{12.0pt}\selectfont -}} & 
\multicolumn{1}{|p{1.74in}|}{\cellcolor[HTML]{FFFFFF}  $Sh\!=\!0.017Ra$ } \\
\hhline{------}
%row no:3
\multicolumn{1}{|p{0.82in}}{\cellcolor[HTML]{FFFFFF}\Centering {\fontsize{10pt}{12.0pt}\selectfont \citep{neufeld2010convective}}} & 
\multicolumn{1}{|p{0.59in}}{\cellcolor[HTML]{FFFFFF}\Centering {\fontsize{10pt}{12.0pt}\selectfont MEG\textsuperscript{$\ast$ }-water}} & 
\multicolumn{1}{|p{0.65in}}{\cellcolor[HTML]{FFFFFF}\Centering {\fontsize{10pt}{12.0pt}\selectfont \textit{Quasi-steady}}} & 
\multicolumn{1}{|p{0.81in}}{\cellcolor[HTML]{FFFFFF}\Centering {\fontsize{10pt}{12.0pt}\selectfont Experimental} \par \Centering {\fontsize{10pt}{12.0pt}\selectfont Numerical}} & 
\multicolumn{1}{|p{0.75in}}{\cellcolor[HTML]{FFFFFF} \( 2\times10^{3}- \)  \par  \( 6\times10^{5} \) } & 
\multicolumn{1}{|p{1.74in}|}{\cellcolor[HTML]{FFFFFF}  \( Sh=0.12Ra^{0.84} \) } \\
\hhline{------}
%row no:4
\multicolumn{1}{|p{0.82in}}{\cellcolor[HTML]{FFFFFF}\Centering {\fontsize{10pt}{12.0pt}\selectfont \citep{pau2010high}}} & 
\multicolumn{1}{|p{0.59in}}{\cellcolor[HTML]{FFFFFF}\Centering {\fontsize{10pt}{12.0pt}\selectfont -}} & 
\multicolumn{1}{|p{0.65in}}{\cellcolor[HTML]{FFFFFF}\Centering {\fontsize{10pt}{12.0pt}\selectfont \textit{Quasi-steady}}} & 
\multicolumn{1}{|p{0.81in}}{\cellcolor[HTML]{FFFFFF}\Centering {\fontsize{10pt}{12.0pt}\selectfont Numerical}} & 
\multicolumn{1}{|p{0.75in}}{\cellcolor[HTML]{FFFFFF}\Centering {\fontsize{10pt}{12.0pt}\selectfont -}} & 
\multicolumn{1}{|p{1.74in}|}{\cellcolor[HTML]{FFFFFF}  \( F=0.017\frac{c_{0}k \Delta  \rho g}{ \mu } \) } \\
\hhline{------}
%row no:5
\multicolumn{1}{|p{0.82in}}{\cellcolor[HTML]{FFFFFF}\Centering {\fontsize{10pt}{12.0pt}\selectfont \citep{backhaus2011convective}}} & 
\multicolumn{1}{|p{0.59in}}{\cellcolor[HTML]{FFFFFF}\Centering {\fontsize{10pt}{12.0pt}\selectfont PPG\textsuperscript{$\ast$ $\ast$ }-water}} & 
\multicolumn{1}{|p{0.65in}}{\cellcolor[HTML]{FFFFFF}\Centering {\fontsize{10pt}{12.0pt}\selectfont \textit{Quasi-steady}}} & 
\multicolumn{1}{|p{0.81in}}{\cellcolor[HTML]{FFFFFF}\Centering {\fontsize{10pt}{12.0pt}\selectfont Experimental}} & 
\multicolumn{1}{|p{0.75in}}{\cellcolor[HTML]{FFFFFF}  \( 6\times10^{3}- \)  \par   \( 9\times10^{4} \) } & 
\multicolumn{1}{|p{1.74in}|}{\cellcolor[HTML]{FFFFFF}  \( Sh=0.045Ra^{0.76} \) } \\
\hhline{------}
%row no:6
\multicolumn{1}{|p{0.82in}}{\cellcolor[HTML]{FFFFFF}\Centering {\fontsize{10pt}{12.0pt}\selectfont \citep{elenius2012time}}} & 
\multicolumn{1}{|p{0.59in}}{\cellcolor[HTML]{FFFFFF}\Centering {\fontsize{10pt}{12.0pt}\selectfont -}} & 
\multicolumn{1}{|p{0.65in}}{\cellcolor[HTML]{FFFFFF}\Centering {\fontsize{10pt}{12.0pt}\selectfont \textit{Quasi-steady}}} & 
\multicolumn{1}{|p{0.81in}}{\cellcolor[HTML]{FFFFFF}\Centering {\fontsize{10pt}{12.0pt}\selectfont Numerical}} & 
\multicolumn{1}{|p{0.75in}}{\cellcolor[HTML]{FFFFFF}\Centering {\fontsize{10pt}{12.0pt}\selectfont -}} & 
\multicolumn{1}{|p{1.74in}|}{\cellcolor[HTML]{FFFFFF}  \( F=0.02\frac{c_{0}k \Delta  \rho g}{ \mu } \) } \\
\hhline{------}
%row no:7
\multicolumn{1}{|p{0.82in}}{\cellcolor[HTML]{FFFFFF}\Centering {\fontsize{10pt}{12.0pt}\selectfont \citep{hidalgo2012scaling}}} & 
\multicolumn{1}{|p{0.59in}}{\cellcolor[HTML]{FFFFFF}\Centering {\fontsize{10pt}{12.0pt}\selectfont -}} & 
\multicolumn{1}{|p{0.65in}}{\cellcolor[HTML]{FFFFFF}\Centering {\fontsize{10pt}{12.0pt}\selectfont \textit{Quasi-steady}}} & 
\multicolumn{1}{|p{0.81in}}{\cellcolor[HTML]{FFFFFF}\Centering {\fontsize{10pt}{12.0pt}\selectfont Numerical}} & 
\multicolumn{1}{|p{0.75in}}{\cellcolor[HTML]{FFFFFF}  \( 5\times10^{3}- \)  \par   \( 3\times10^{4} \) } & 
\multicolumn{1}{|p{1.74in}|}{\cellcolor[HTML]{FFFFFF}  \( F=a\frac{c_{0}k \Delta  \rho g}{ \mu } \) } \\
\hhline{------}
%row no:8
\multicolumn{1}{|p{0.82in}}{\cellcolor[HTML]{FFFFFF}\Centering {\fontsize{10pt}{12.0pt}\selectfont \citep{farajzadeh2013empirical}}} & 
\multicolumn{1}{|p{0.59in}}{\cellcolor[HTML]{FFFFFF}\Centering {\fontsize{10pt}{12.0pt}\selectfont -}} & 
\multicolumn{1}{|p{0.65in}}{\cellcolor[HTML]{FFFFFF}\Centering {\fontsize{10pt}{12.0pt}\selectfont \textit{Quasi-steady}}} & 
\multicolumn{1}{|p{0.81in}}{\cellcolor[HTML]{FFFFFF}\Centering {\fontsize{10pt}{12.0pt}\selectfont Numerical}} & 
\multicolumn{1}{|p{0.75in}}{\cellcolor[HTML]{FFFFFF}  \( 1\times10^{3}- \)  \par   \( 8\times10^{3} \) } & 
\multicolumn{1}{|p{1.74in}|}{\cellcolor[HTML]{FFFFFF}  \( Sh=0.0794Ra^{0.832} \) } \\
\hhline{------}
%row no:9
\multicolumn{1}{|p{0.82in}}{\cellcolor[HTML]{FFFFFF}\Centering {\fontsize{10pt}{12.0pt}\selectfont \citep{tsai2013density}}} & 
\multicolumn{1}{|p{0.59in}}{\cellcolor[HTML]{FFFFFF}\Centering {\fontsize{10pt}{12.0pt}\selectfont PPG\textsuperscript{$\ast$ $\ast$ }-water}} & 
\multicolumn{1}{|p{0.65in}}{\cellcolor[HTML]{FFFFFF}\Centering {\fontsize{10pt}{12.0pt}\selectfont \textit{Quasi-steady}}} & 
\multicolumn{1}{|p{0.81in}}{\cellcolor[HTML]{FFFFFF}\Centering {\fontsize{10pt}{12.0pt}\selectfont Experimental}} & 
\multicolumn{1}{|p{0.75in}}{\cellcolor[HTML]{FFFFFF}  \( 5\times10^{3}- \)  \par   \( 1\times10^{5} \) } & 
\multicolumn{1}{|p{1.74in}|}{\cellcolor[HTML]{FFFFFF}  \( Sh=0.037Ra^{0.84} \) } \\
\hhline{------}
%row no:10
\multicolumn{1}{|p{0.82in}}{\cellcolor[HTML]{FFFFFF}\Centering {\fontsize{10pt}{12.0pt}\selectfont \citep{elenius2014convective}}} & 
\multicolumn{1}{|p{0.59in}}{\cellcolor[HTML]{FFFFFF}\Centering {\fontsize{10pt}{12.0pt}\selectfont -}} & 
\multicolumn{1}{|p{0.65in}}{\cellcolor[HTML]{FFFFFF}\Centering {\fontsize{10pt}{12.0pt}\selectfont \textit{Quasi-steady}}} & 
\multicolumn{1}{|p{0.81in}}{\cellcolor[HTML]{FFFFFF}\Centering {\fontsize{10pt}{12.0pt}\selectfont Numerical}} & 
\multicolumn{1}{|p{0.75in}}{\cellcolor[HTML]{FFFFFF}\Centering {\fontsize{10pt}{12.0pt}\selectfont -}} & 
\multicolumn{1}{|p{1.74in}|}{\cellcolor[HTML]{FFFFFF}  \( F=0.021 \frac{c_{0}k \Delta  \rho g}{ \mu } \) } \\
\hhline{------}
%row no:11
\multicolumn{1}{|p{0.82in}}{\cellcolor[HTML]{FFFFFF}\Centering {\fontsize{10pt}{12.0pt}\selectfont \citep{green2014steady}}} & 
\multicolumn{1}{|p{0.59in}}{\cellcolor[HTML]{FFFFFF}\Centering {\fontsize{10pt}{12.0pt}\selectfont -}} & 
\multicolumn{1}{|p{0.65in}}{\cellcolor[HTML]{FFFFFF}\Centering {\fontsize{10pt}{12.0pt}\selectfont \textit{Quasi-steady}}} & 
\multicolumn{1}{|p{0.81in}}{\cellcolor[HTML]{FFFFFF}\Centering {\fontsize{10pt}{12.0pt}\selectfont Numerical}} & 
\multicolumn{1}{|p{0.75in}}{\cellcolor[HTML]{FFFFFF}\Centering {\fontsize{10pt}{12.0pt}\selectfont -}} & 
\multicolumn{1}{|p{1.74in}|}{\cellcolor[HTML]{FFFFFF}  \( F=0.017\sqrt[]{k_{v}k_{h}}\frac{c_{0} \Delta  \rho g}{ \mu } \) } \\
\hhline{------}
%row no:12
\multicolumn{1}{|p{0.82in}}{\cellcolor[HTML]{FFFFFF}\Centering {\fontsize{10pt}{12.0pt}\selectfont \citep{mojtaba2014experimental}}} & 
\multicolumn{1}{|p{0.59in}}{\cellcolor[HTML]{FFFFFF}\Centering {\fontsize{10pt}{12.0pt}\selectfont CO$_2$-brine (NaCl)}} & 
\multicolumn{1}{|p{0.65in}}{\cellcolor[HTML]{FFFFFF}\Centering {\fontsize{10pt}{12.0pt}\selectfont \textit{Quasi-steady}}} & 
\multicolumn{1}{|p{0.81in}}{\cellcolor[HTML]{FFFFFF}\Centering {\fontsize{10pt}{12.0pt}\selectfont Experimental}} & 
\multicolumn{1}{|p{0.75in}}{\cellcolor[HTML]{FFFFFF}\Centering {\fontsize{10pt}{12.0pt}\selectfont 182-20860}} & 
\multicolumn{1}{|p{1.74in}|}{\cellcolor[HTML]{FFFFFF}  \( Sh=0.0228Ra^{0.7897} \) } \\
\hhline{------}
%row no:13
\multicolumn{1}{|p{0.82in}}{\multirow{1}{*}{\begin{tabular}{p{0.82in}}\cellcolor[HTML]{FFFFFF}\Centering {\fontsize{10pt}{12.0pt}\selectfont \citep{slim2014solutal}}\\\end{tabular}}} & 
\multicolumn{1}{|p{0.59in}}{\multirow{1}{*}{\begin{tabular}{p{0.59in}}\cellcolor[HTML]{FFFFFF}\Centering {\fontsize{10pt}{12.0pt}\selectfont -}\\\end{tabular}}} & 
\multicolumn{1}{|p{0.65in}}{\cellcolor[HTML]{FFFFFF}\Centering {\fontsize{10pt}{12.0pt}\selectfont \textit{Quasi-steady}}} & 
\multicolumn{1}{|p{0.81in}}{\multirow{1}{*}{\begin{tabular}{p{0.81in}}\cellcolor[HTML]{FFFFFF}\Centering {\fontsize{10pt}{12.0pt}\selectfont Numerical}\\\end{tabular}}} & 
\multicolumn{1}{|p{0.75in}}{\multirowcell{2}{}{\begin{tabular}{p{0.75in}}\cellcolor[HTML]{FFFFFF}  \( 2\times10^{3}- \) \\  \( 5\times10^{5} \) \\\end{tabular}}} & 
\multicolumn{1}{|p{1.74in}|}{\cellcolor[HTML]{FFFFFF}  \( F=0.017\frac{c_{0}k \Delta  \rho g}{ \mu } \) } \\
\hhline{~~-~~-}
%row no:14
\multicolumn{1}{|p{0.82in}}{\cellcolor[HTML]{FFFFFF}} & 
\multicolumn{1}{|p{0.59in}}{\cellcolor[HTML]{FFFFFF}} & 
\multicolumn{1}{|p{0.65in}}{\cellcolor[HTML]{FFFFFF}\Centering {\fontsize{10pt}{12.0pt}\selectfont \textit{Shut-down}}} & 
\multicolumn{1}{|p{0.81in}}{\cellcolor[HTML]{FFFFFF}} & 
\multicolumn{1}{|p{0.75in}}{\cellcolor[HTML]{FFFFFF}} & 
\multicolumn{1}{|p{1.74in}|}{\cellcolor[HTML]{FFFFFF}  \( Sh=\frac{16.8}{ \left[ 0.73 \left( \frac{t}{Ra}-16 \right) +31.5 \right] ^{2}} \) } \\
\hhline{------}
%row no:15
\multicolumn{1}{|p{0.82in}}{\cellcolor[HTML]{FFFFFF}\Centering {\fontsize{10pt}{12.0pt}\selectfont \citep{martinez2016two}}} & 
\multicolumn{1}{|p{0.59in}}{\cellcolor[HTML]{FFFFFF}\Centering {\fontsize{10pt}{12.0pt}\selectfont -}} & 
\multicolumn{1}{|p{0.65in}}{\cellcolor[HTML]{FFFFFF}\Centering {\fontsize{10pt}{12.0pt}\selectfont \textit{Quasi-steady}}} & 
\multicolumn{1}{|p{0.81in}}{\cellcolor[HTML]{FFFFFF}\Centering {\fontsize{10pt}{12.0pt}\selectfont Numerical}} & 
\multicolumn{1}{|p{0.75in}}{\cellcolor[HTML]{FFFFFF}\Centering {\fontsize{10pt}{12.0pt}\selectfont -}} & 
\multicolumn{1}{|p{1.74in}|}{\cellcolor[HTML]{FFFFFF}  \( F= \left( 0.018, 0.019 \right) \frac{c_{0}k \Delta  \rho g}{ \mu } \) } \\
\hhline{------}
%row no:16
\multicolumn{1}{|p{0.82in}}{\cellcolor[HTML]{FFFFFF}\Centering {\fontsize{10pt}{12.0pt}\selectfont \citep{wang2016three}}} & 
\multicolumn{1}{|p{0.59in}}{\cellcolor[HTML]{FFFFFF}\Centering {\fontsize{10pt}{12.0pt}\selectfont NaCl solution- MEG}} & 
\multicolumn{1}{|p{0.65in}}{\cellcolor[HTML]{FFFFFF}\Centering {\fontsize{10pt}{12.0pt}\selectfont \textit{Quasi-steady}}} & 
\multicolumn{1}{|p{0.81in}}{\cellcolor[HTML]{FFFFFF}\Centering {\fontsize{10pt}{12.0pt}\selectfont Experimental}} & 
\multicolumn{1}{|p{0.75in}}{\cellcolor[HTML]{FFFFFF}\Centering {\fontsize{10pt}{12.0pt}\selectfont 2600-16036}} & 
\multicolumn{1}{|p{1.74in}|}{\cellcolor[HTML]{FFFFFF}  \( Sh=0.13Ra^{0.93} \) } \\
\hhline{------}
%row no:17
\multicolumn{1}{|p{0.82in}}{\multirow{1}{*}{\begin{tabular}{p{0.82in}}\cellcolor[HTML]{FFFFFF}\Centering {\fontsize{10pt}{12.0pt}\selectfont \citep{de2017dissolution}}\\\end{tabular}}} & 
\multicolumn{1}{|p{0.59in}}{\multirow{1}{*}{\begin{tabular}{p{0.59in}}\cellcolor[HTML]{FFFFFF}\Centering {\fontsize{10pt}{12.0pt}\selectfont -}\\\cellcolor[HTML]{FFFFFF}\end{tabular}}} & 
\multicolumn{1}{|p{0.65in}}{\cellcolor[HTML]{FFFFFF}\Centering {\fontsize{10pt}{12.0pt}\selectfont \textit{Quasi-steady}}} & 
\multicolumn{1}{|p{0.81in}}{\multirow{3}{*}{\begin{tabular}{p{0.81in}}\cellcolor[HTML]{FFFFFF}\Centering {\fontsize{10pt}{12.0pt}\selectfont Numerical}\\\cellcolor[HTML]{FFFFFF}\end{tabular}}} & 
\multicolumn{1}{|p{0.75in}}{\multirow{1}{*}{\begin{tabular}{p{0.75in}}\cellcolor[HTML]{FFFFFF}  \( 1\times10^{3}- \) \\ \cellcolor[HTML]{FFFFFF} \( 2\times10^{4} \)    \\\end{tabular}}} &  
\multicolumn{1}{|p{1.74in}|}{\cellcolor[HTML]{FFFFFF}  \( F=0.017\sqrt[]{k_{v}k_{h}}\frac{c_{0} \Delta  \rho g}{ \mu } \) } \\
\hhline{~~-~~-}
%row no:18
\multicolumn{1}{|p{0.82in}}{\cellcolor[HTML]{FFFFFF}} & 
\multicolumn{1}{|p{0.59in}}{\cellcolor[HTML]{FFFFFF}} & 
\multicolumn{1}{|p{0.65in}}{\cellcolor[HTML]{FFFFFF}\Centering {\fontsize{10pt}{12.0pt}\selectfont \textit{Shut-down}}} & 
\multicolumn{1}{|p{0.81in}}{\cellcolor[HTML]{FFFFFF}} & 
\multicolumn{1}{|p{0.75in}}{\cellcolor[HTML]{FFFFFF}} & 
\multicolumn{1}{|p{1.74in}|}{\cellcolor[HTML]{FFFFFF}  \( Sh=\frac{4a \left( \frac{k_{v}}{k_{h}} \right) ^{a'}}{ \left( 1+4a \left( \frac{k_{v}}{k_{h}} \right) ^{a'}t/Ra \right) ^{2}} \) } \\
\hhline{------}
%row no:19
\multicolumn{1}{|p{0.82in}}{\cellcolor[HTML]{FFFFFF}\Centering {\fontsize{10pt}{12.0pt}\selectfont \citep{taheri2017qualitative}}} & 
\multicolumn{1}{|p{0.59in}}{\cellcolor[HTML]{FFFFFF}\Centering {\fontsize{10pt}{12.0pt}\selectfont CO$_2$-water}} & 
\multicolumn{1}{|p{0.65in}}{\cellcolor[HTML]{FFFFFF}\Centering {\fontsize{10pt}{12.0pt}\selectfont \textit{Quasi-steady}}} & 
\multicolumn{1}{|p{0.81in}}{\cellcolor[HTML]{FFFFFF}\Centering {\fontsize{10pt}{12.0pt}\selectfont Experimental}} & 
\multicolumn{1}{|p{0.75in}}{\cellcolor[HTML]{FFFFFF}\Centering {\fontsize{10pt}{12.0pt}\selectfont 709- 9627}} & 
\multicolumn{1}{|p{1.74in}|}{\cellcolor[HTML]{FFFFFF}  \( F=\frac{0.021 \Delta  \rho g.cos \left(  \theta  \right) kc_{0}}{ \mu } \) } \\
\hhline{------}
%row no:20
\multicolumn{1}{|p{0.82in}}{\multirow{1}{*}{\begin{tabular}{p{0.82in}}\cellcolor[HTML]{FFFFFF}\Centering {\fontsize{10pt}{12.0pt}\selectfont \citep{newell2018experimental}}\\\cellcolor[HTML]{FFFFFF}\end{tabular}}} & 
\multicolumn{1}{|p{0.59in}}{\multirow{1}{*}{\begin{tabular}{p{0.59in}}\cellcolor[HTML]{FFFFFF}\Centering {\fontsize{10pt}{12.0pt}\selectfont CO$_2$-water}\\\cellcolor[HTML]{FFFFFF}\end{tabular}}} & 
\multicolumn{1}{|p{0.65in}}{\cellcolor[HTML]{FFFFFF}\Centering {\fontsize{10pt}{12.0pt}\selectfont \textit{Flux-growth}}} & 
\multicolumn{1}{|p{0.81in}}{\multirowcell{1}{}{\begin{tabular}{p{0.81in}}\cellcolor[HTML]{FFFFFF}\Centering {\fontsize{10pt}{12.0pt}\selectfont Experimental}\\\cellcolor[HTML]{FFFFFF}\Centering {\fontsize{10pt}{12.0pt}\selectfont Numerical}\end{tabular}}} & 
\multicolumn{1}{|p{0.75in}}{\multirow{3}{*}{\begin{tabular}{p{0.75in}}\cellcolor[HTML]{FFFFFF}\Centering {\fontsize{10pt}{12.0pt}\selectfont 2093-16256}\\\cellcolor[HTML]{FFFFFF}\end{tabular}}} & 
\multicolumn{1}{|p{1.74in}|}{\cellcolor[HTML]{FFFFFF}  \( F=a't^{0.5} \) } \\
\hhline{~~-~~-}
%row no:21
\multicolumn{1}{|p{0.82in}}{\cellcolor[HTML]{FFFFFF}} & 
\multicolumn{1}{|p{0.59in}}{\cellcolor[HTML]{FFFFFF}} & 
\multicolumn{1}{|p{0.65in}}{\cellcolor[HTML]{FFFFFF}\Centering {\fontsize{10pt}{12.0pt}\selectfont \textit{Quasi-steady}}} & 
\multicolumn{1}{|p{0.81in}}{\cellcolor[HTML]{FFFFFF}} & 
\multicolumn{1}{|p{0.75in}}{\cellcolor[HTML]{FFFFFF}} & 
\multicolumn{1}{|p{1.74in}|}{\cellcolor[HTML]{FFFFFF}  \( F=a\frac{c_{0}k \Delta  \rho g}{ \mu  \varphi } \) } \\
\hhline{~~-~~-}
%row no:22
\multicolumn{1}{|p{0.82in}}{\cellcolor[HTML]{FFFFFF}} & 
\multicolumn{1}{|p{0.59in}}{\cellcolor[HTML]{FFFFFF}} & 
\multicolumn{1}{|p{0.65in}}{\cellcolor[HTML]{FFFFFF}\Centering {\fontsize{10pt}{12.0pt}\selectfont \textit{Shut-down}}} & 
\multicolumn{1}{|p{0.81in}}{\cellcolor[HTML]{FFFFFF}} & 
\multicolumn{1}{|p{0.75in}}{\cellcolor[HTML]{FFFFFF}} & 
\multicolumn{1}{|p{1.74in}|}{\cellcolor[HTML]{FFFFFF}  \( F=a^{'}t^{-1.75} \) } \\
\hhline{------}
%row no:23
\multicolumn{1}{|p{0.82in}}{\multirow{1}{*}{\begin{tabular}{p{0.82in}}\cellcolor[HTML]{FFFFFF}\Centering {\fontsize{10pt}{12.0pt}\selectfont \citep{wen2018dynamics}}\\\end{tabular}}} & 
\multicolumn{1}{|p{0.59in}}{\multirow{1}{*}{\begin{tabular}{p{0.59in}}\cellcolor[HTML]{FFFFFF}\Centering {\fontsize{10pt}{12.0pt}\selectfont -}\\\end{tabular}}} & 
\multicolumn{1}{|p{0.65in}}{\cellcolor[HTML]{FFFFFF}\Centering {\fontsize{10pt}{12.0pt}\selectfont \textit{Quasi-steady}}} & 
\multicolumn{1}{|p{0.81in}}{\multirow{1}{*}{\begin{tabular}{p{0.81in}}\cellcolor[HTML]{FFFFFF}\Centering {\fontsize{10pt}{12.0pt}\selectfont Numerical}\\\end{tabular}}} & 
\multicolumn{1}{|p{0.75in}}{\multirow{1}{*}{\begin{tabular}{p{0.75in}}\cellcolor[HTML]{FFFFFF}  \( 1\times10^{4}- \) \\ { \( 5\times10^{4} \) }\\\end{tabular}}} & 
\multicolumn{1}{|p{1.74in}|}{\cellcolor[HTML]{FFFFFF}  \( Sh=\frac{0.0168 Ra_{0}}{ \left( 0.0168a t_{a}+1 \right) ^{2}} \) } \\
\hhline{~~-~~-}
%row no:24
\multicolumn{1}{|p{0.82in}}{\cellcolor[HTML]{FFFFFF}} & 
\multicolumn{1}{|p{0.59in}}{\cellcolor[HTML]{FFFFFF}} & 
\multicolumn{1}{|p{0.65in}}{\cellcolor[HTML]{FFFFFF}\Centering {\fontsize{10pt}{12.0pt}\selectfont \textit{Shut-down}}} & 
\multicolumn{1}{|p{0.81in}}{\cellcolor[HTML]{FFFFFF}} & 
\multicolumn{1}{|p{0.75in}}{\cellcolor[HTML]{FFFFFF}} & 
\multicolumn{1}{|p{1.74in}|}{\cellcolor[HTML]{FFFFFF}  \( Sh=\frac{0.0317Ra_{0}}{ \left( 0.0317 \left( 1+a \right) t_{a}+0.861 \right) ^{2}} \) } \\
\hhline{------}

\end{tabular}
\end{table*}}

%%%%%%%%%%%%%%%%%%%% Table No: 3 starts here %%%%%%%%%%%%%%%%%%%%

\begin{table*}
 			\centering
			\caption{The details of experimental tests with their order number, brine composition, permeability, Rayleigh value, initial pressure, and diffusion coefficient. The details of  correlations that are used to calculate  Ra are provided in  Appendix A.}
\begin{tabular}{p{0.52in}p{1.62in}p{0.85in}p{0.57in}p{0.95in}p{0.78in}}
\hline
%row no:1
\multicolumn{1}{|p{0.52in}}{\Centering Test  \par \Centering Number} & 
\multicolumn{1}{|p{1.62in}}{\Centering brine composition \par \Centering (mole based)} & 
\multicolumn{1}{|p{0.85in}}{\Centering Permeability \par \Centering (\textit{D})} & 
\multicolumn{1}{|p{0.57in}}{\Centering Rayleigh \par \Centering number} & 
\multicolumn{1}{|p{0.95in}}{\Centering Initial Pressure \par \Centering ({psi})} & 
\multicolumn{1}{|p{0.78in}|}{\Centering Diffusion \par \Centering$\times$\( 10^{9} \left( \frac{m^{2}}{s} \right)  \) } \\
\hhline{------}
%row no:2
\multicolumn{1}{|p{0.52in}}{\Centering 1} & 
\multicolumn{1}{|p{1.62in}}{\Centering 2 NaCl} & 
\multicolumn{1}{|p{0.85in}}{\Centering 550} & 
\multicolumn{1}{|p{0.57in}}{\Centering {\fontsize{10pt}{12.0pt}\selectfont 4444}} & 
\multicolumn{1}{|p{0.95in}}{\Centering 523.1} & 
\multicolumn{1}{|p{0.78in}|}{\Centering 3.8} \\
\hhline{------}
%row no:3
\multicolumn{1}{|p{0.52in}}{\Centering 2} & 
\multicolumn{1}{|p{1.62in}}{\Centering 2 NaCl} & 
\multicolumn{1}{|p{0.85in}}{\Centering 400} & 
\multicolumn{1}{|p{0.57in}}{\Centering {\fontsize{10pt}{12.0pt}\selectfont 3272}} & 
\multicolumn{1}{|p{0.95in}}{\Centering 511.2} & 
\multicolumn{1}{|p{0.78in}|}{\Centering 3.7} \\
\hhline{------}
%row no:4
\multicolumn{1}{|p{0.52in}}{\Centering 3} & 
\multicolumn{1}{|p{1.62in}}{\Centering 1 NaCl} & 
\multicolumn{1}{|p{0.85in}}{\Centering 550} & 
\multicolumn{1}{|p{0.57in}}{\Centering {\fontsize{10pt}{12.0pt}\selectfont 4841}} & 
\multicolumn{1}{|p{0.95in}}{\Centering 535.3} & 
\multicolumn{1}{|p{0.78in}|}{\Centering 5} \\
\hhline{------}
%row no:5
\multicolumn{1}{|p{0.52in}}{\Centering 4} & 
\multicolumn{1}{|p{1.62in}}{\Centering 1 NaCl} & 
\multicolumn{1}{|p{0.85in}}{\Centering 400} & 
\multicolumn{1}{|p{0.57in}}{\Centering {\fontsize{10pt}{12.0pt}\selectfont 3514}} & 
\multicolumn{1}{|p{0.95in}}{\Centering 510.6} & 
\multicolumn{1}{|p{0.78in}|}{\Centering 4.9} \\
\hhline{------}
%row no:6
\multicolumn{1}{|p{0.52in}}{\Centering 5} & 
\multicolumn{1}{|p{1.62in}}{\Centering 1.6 {NaCl}+ 0.2 CaCl$_2$} & 
\multicolumn{1}{|p{0.85in}}{\Centering 550} & 
\multicolumn{1}{|p{0.57in}}{\Centering {\fontsize{10pt}{12.0pt}\selectfont 3893}} & 
\multicolumn{1}{|p{0.95in}}{\Centering 512.7} & 
\multicolumn{1}{|p{0.78in}|}{\Centering 4.2} \\
\hhline{------}
%row no:7
\multicolumn{1}{|p{0.52in}}{\Centering 6} & 
\multicolumn{1}{|p{1.62in}}{\Centering 1.6 NaCl + 0.2 CaCl$_2$} & 
\multicolumn{1}{|p{0.85in}}{\Centering 400} & 
\multicolumn{1}{|p{0.57in}}{\Centering {\fontsize{10pt}{12.0pt}\selectfont 2919}} & 
\multicolumn{1}{|p{0.95in}}{\Centering 514.7} & 
\multicolumn{1}{|p{0.78in}|}{\Centering 4.1} \\
\hhline{------}
%row no:8
\multicolumn{1}{|p{0.52in}}{\Centering 7} & 
\multicolumn{1}{|p{1.62in}}{\Centering 0.8 NaCl + 0.1 CaCl$_2$} & 
\multicolumn{1}{|p{0.85in}}{\Centering 550} & 
\multicolumn{1}{|p{0.57in}}{\Centering {\fontsize{10pt}{12.0pt}\selectfont 4283}} & 
\multicolumn{1}{|p{0.95in}}{\Centering 502.6} & 
\multicolumn{1}{|p{0.78in}|}{\Centering 5.3} \\
\hhline{------}
%row no:9
\multicolumn{1}{|p{0.52in}}{\Centering 8} & 
\multicolumn{1}{|p{1.62in}}{\Centering 0.8 NaCl + 0.1 CaCl$_2$} & 
\multicolumn{1}{|p{0.85in}}{\Centering 400} & 
\multicolumn{1}{|p{0.57in}}{\Centering {\fontsize{10pt}{12.0pt}\selectfont 3242}} & 
\multicolumn{1}{|p{0.95in}}{\Centering 505.5} & 
\multicolumn{1}{|p{0.78in}|}{\Centering 5.2} \\
\hhline{------}

\end{tabular}
 \end{table*}

%%%%%%%%%%%%%%%%%%%% Table No: 3 ends here %%%%%%%%%%%%%%%%%%%%

%%%%%%%%%%%%%%%%%%%% Table No: 7 starts here %%%%%%%%%%%%%%%%%%%%

\begin{table*}
 			\centering
			\caption{Fitted and estimated factors for the modified dissolution flux equation (after \citep{wen2018dynamics})  and transition times between different   dissolution regimes}
\begin{tabular}{p{0.22in}p{0.13in}p{0.23in}p{0.23in}p{0.24in}p{0.23in}p{0.26in}p{0.27in}p{0.23in}p{0.23in}p{0.26in}p{0.33in}p{0.23in}}
\hline
%row no:1
\multicolumn{1}{|p{0.27in}}{\Centering {\fontsize{10pt}{12.0pt}\selectfont Case}} & 
\multicolumn{1}{|p{0.13in}}{\Centering { \color{purple}\( \frac{F_{c}}{C_{s}^{2}} \)} } & 
\multicolumn{1}{|p{0.26in}}{\Centering  { \color[HTML]{002060}\( m_{fit.}^{F.G} \)  \par \Centering  \( \times10^{4} \) } }& 
\multicolumn{1}{|p{0.23in}}{\Centering  { \color[HTML]{002060}\( b_{fit.}^{F.G} \)  \par \Centering  \( \times10^{6} \) } }& 
\multicolumn{1}{|p{0.26in}}{\Centering  { \color[HTML]{002060}\( m_{est.}^{F.G} \)  \par \Centering  \( \times10^{4} \) } }& 
\multicolumn{1}{|p{0.23in}}{\Centering  { \color[HTML]{002060}\( b_{est.}^{F.G} \)  \par \Centering  \( \times10^{6} \) } }& 
\multicolumn{1}{|p{0.26in}}{\Centering { \color{blue} \( t_{D,fit.}^{Q.S} \)  \par \Centering  \( \times10^{4} \) } }& 
\multicolumn{1}{|p{0.27in}}{\Centering  { \color{blue}\( t_{D,est.}^{Q.S} \)  \par \Centering  \( \times10^{4} \) }} & 
\multicolumn{1}{|p{0.25in}}{\Centering  { \color{teal}\( a_{fit.}^{Q.S} \)  \par \Centering  \( \times10^{4} \) } }& 
\multicolumn{1}{|p{0.25in}}{\Centering  { \color{teal}\( a_{est.}^{Q.S} \)  \par \Centering  \( \times10^{4} \) } }& 
\multicolumn{1}{|p{0.26in}}{\Centering {\color[HTML]{0070C0} \( t_{D,fit.}^{Sh} \)  \par \Centering  \( \times10^{4} \) } }& 
\multicolumn{1}{|p{0.33in}}{\Centering  { \color[HTML]{0070C0}\( t_{D,est.}^{Sh} \)  \par \Centering  \(\times10^{4} \) } }& 
\multicolumn{1}{|p{0.25in}|}{\Centering  { \color{cyan}\( a_{fit.}^{S.D} \)  \par \Centering  \( \times10^{9} \) }} \\
\hhline{-------------}
%row no:2
\multicolumn{1}{|p{0.22in}}{\Centering {\fontsize{10pt}{12.0pt}\selectfont 1}} & 
\multicolumn{1}{|p{0.16in}}{\Centering {\fontsize{10pt}{12.0pt}\selectfont 7}} & 
\multicolumn{1}{|p{0.23in}}{\Centering {\fontsize{10pt}{12.0pt}\selectfont 77}} & 
\multicolumn{1}{|p{0.27in}}{\Centering {\fontsize{10pt}{12.0pt}\selectfont -124}} & 
\multicolumn{1}{|p{0.24in}}{\Centering {\fontsize{10pt}{12.0pt}\selectfont 78}} & 
\multicolumn{1}{|p{0.27in}}{\Centering {\fontsize{10pt}{12.0pt}\selectfont -134}} & 
\multicolumn{1}{|p{0.26in}}{\Centering {\fontsize{10pt}{12.0pt}\selectfont 28.73}} & 
\multicolumn{1}{|p{0.3in}}{\Centering {\fontsize{10pt}{12.0pt}\selectfont 28.58}} & 
\multicolumn{1}{|p{0.23in}}{\Centering {\fontsize{10pt}{12.0pt}\selectfont 2.90}} & 
\multicolumn{1}{|p{0.23in}}{\Centering {\fontsize{10pt}{12.0pt}\selectfont 2.95}} & 
\multicolumn{1}{|p{0.29in}}{\Centering {\fontsize{10pt}{12.0pt}\selectfont 54.13}} & 
\multicolumn{1}{|p{0.33in}}{\Centering {\fontsize{10pt}{12.0pt}\selectfont 53.03}} & 
\multicolumn{1}{|p{0.23in}|}{\Centering {\fontsize{10pt}{12.0pt}\selectfont 8.3}} \\
\hhline{-------------}
%row no:3
\multicolumn{1}{|p{0.22in}}{\cellcolor[HTML]{FFFFFF}\Centering {\fontsize{10pt}{12.0pt}\selectfont 2}} & 
\multicolumn{1}{|p{0.13in}}{\cellcolor[HTML]{FFFFFF}\Centering {\fontsize{10pt}{12.0pt}\selectfont 5}} & 
\multicolumn{1}{|p{0.23in}}{\cellcolor[HTML]{FFFFFF}\Centering {\fontsize{10pt}{12.0pt}\selectfont 45}} & 
\multicolumn{1}{|p{0.23in}}{\cellcolor[HTML]{FFFFFF}\Centering {\fontsize{10pt}{12.0pt}\selectfont -52}} & 
\multicolumn{1}{|p{0.24in}}{\cellcolor[HTML]{FFFFFF}\Centering {\fontsize{10pt}{12.0pt}\selectfont 43}} & 
\multicolumn{1}{|p{0.23in}}{\cellcolor[HTML]{FFFFFF}\Centering {\fontsize{10pt}{12.0pt}\selectfont -50}} & 
\multicolumn{1}{|p{0.3in}}{\cellcolor[HTML]{FFFFFF}\Centering {\fontsize{10pt}{12.0pt}\selectfont 38.47}} & 
\multicolumn{1}{|p{0.27in}}{\cellcolor[HTML]{FFFFFF}\Centering {\fontsize{10pt}{12.0pt}\selectfont 37.64}} & 
\multicolumn{1}{|p{0.23in}}{\cellcolor[HTML]{FFFFFF}\Centering {\fontsize{10pt}{12.0pt}\selectfont 2.20}} & 
\multicolumn{1}{|p{0.23in}}{\cellcolor[HTML]{FFFFFF}\Centering {\fontsize{10pt}{12.0pt}\selectfont 2.09}} & 
\multicolumn{1}{|p{0.26in}}{\cellcolor[HTML]{FFFFFF}\Centering {\fontsize{10pt}{12.0pt}\selectfont 71.00}} & 
\multicolumn{1}{|p{0.33in}}{\cellcolor[HTML]{FFFFFF}\Centering {\fontsize{10pt}{12.0pt}\selectfont 70.47}} & 
\multicolumn{1}{|p{0.23in}|}{\cellcolor[HTML]{FFFFFF}\Centering {\fontsize{10pt}{12.0pt}\selectfont 11}} \\
\hhline{-------------}
%row no:4
\multicolumn{1}{|p{0.22in}}{\Centering {\fontsize{10pt}{12.0pt}\selectfont 3}} & 
\multicolumn{1}{|p{0.13in}}{\Centering {\fontsize{10pt}{12.0pt}\selectfont 4.1}} & 
\multicolumn{1}{|p{0.23in}}{\Centering {\fontsize{10pt}{12.0pt}\selectfont 96}} & 
\multicolumn{1}{|p{0.26in}}{\Centering {\fontsize{10pt}{12.0pt}\selectfont -173}} & 
\multicolumn{1}{|p{0.24in}}{\Centering {\fontsize{10pt}{12.0pt}\selectfont 90}} & 
\multicolumn{1}{|p{0.26in}}{\Centering {\fontsize{10pt}{12.0pt}\selectfont -162}} & 
\multicolumn{1}{|p{0.26in}}{\Centering {\fontsize{10pt}{12.0pt}\selectfont 26.47}} & 
\multicolumn{1}{|p{0.27in}}{\Centering {\fontsize{10pt}{12.0pt}\selectfont 26.46}} & 
\multicolumn{1}{|p{0.23in}}{\Centering {\fontsize{10pt}{12.0pt}\selectfont 3.30}} & 
\multicolumn{1}{|p{0.23in}}{\Centering {\fontsize{10pt}{12.0pt}\selectfont 3.24}} & 
\multicolumn{1}{|p{0.26in}}{\Centering {\fontsize{10pt}{12.0pt}\selectfont 48.10}} & 
\multicolumn{1}{|p{0.33in}}{\Centering {\fontsize{10pt}{12.0pt}\selectfont 48.99}} & 
\multicolumn{1}{|p{0.23in}|}{\Centering {\fontsize{10pt}{12.0pt}\selectfont 6.5}} \\
\hhline{-------------}
%row no:5
\multicolumn{1}{|p{0.22in}}{\cellcolor[HTML]{FFFFFF}\Centering {\fontsize{10pt}{12.0pt}\selectfont 4}} & 
\multicolumn{1}{|p{0.13in}}{\cellcolor[HTML]{FFFFFF}\Centering {\fontsize{10pt}{12.0pt}\selectfont 3.4}} & 
\multicolumn{1}{|p{0.23in}}{\cellcolor[HTML]{FFFFFF}\Centering {\fontsize{10pt}{12.0pt}\selectfont 55}} & 
\multicolumn{1}{|p{0.23in}}{\cellcolor[HTML]{FFFFFF}\Centering {\fontsize{10pt}{12.0pt}\selectfont -65}} & 
\multicolumn{1}{|p{0.24in}}{\cellcolor[HTML]{FFFFFF}\Centering {\fontsize{10pt}{12.0pt}\selectfont 50}} & 
\multicolumn{1}{|p{0.23in}}{\cellcolor[HTML]{FFFFFF}\Centering {\fontsize{10pt}{12.0pt}\selectfont -68}} & 
\multicolumn{1}{|p{0.26in}}{\cellcolor[HTML]{FFFFFF}\Centering {\fontsize{10pt}{12.0pt}\selectfont 34.27}} & 
\multicolumn{1}{|p{0.27in}}{\cellcolor[HTML]{FFFFFF}\Centering {\fontsize{10pt}{12.0pt}\selectfont 35.30}} & 
\multicolumn{1}{|p{0.23in}}{\cellcolor[HTML]{FFFFFF}\Centering {\fontsize{10pt}{12.0pt}\selectfont 2.30}} & 
\multicolumn{1}{|p{0.23in}}{\cellcolor[HTML]{FFFFFF}\Centering {\fontsize{10pt}{12.0pt}\selectfont 2.26}} & 
\multicolumn{1}{|p{0.26in}}{\cellcolor[HTML]{FFFFFF}\Centering {\fontsize{10pt}{12.0pt}\selectfont 64.47}} & 
\multicolumn{1}{|p{0.33in}}{\cellcolor[HTML]{FFFFFF}\Centering {\fontsize{10pt}{12.0pt}\selectfont 65.95}} & 
\multicolumn{1}{|p{0.23in}|}{\cellcolor[HTML]{FFFFFF}\Centering {\fontsize{10pt}{12.0pt}\selectfont 5.5}} \\
\hhline{-------------}
%row no:6
\multicolumn{1}{|p{0.22in}}{\Centering {\fontsize{10pt}{12.0pt}\selectfont 5}} & 
\multicolumn{1}{|p{0.13in}}{\Centering {\fontsize{10pt}{12.0pt}\selectfont 4.5}} & 
\multicolumn{1}{|p{0.23in}}{\Centering {\fontsize{10pt}{12.0pt}\selectfont 61}} & 
\multicolumn{1}{|p{0.23in}}{\Centering {\fontsize{10pt}{12.0pt}\selectfont -92}} & 
\multicolumn{1}{|p{0.24in}}{\Centering {\fontsize{10pt}{12.0pt}\selectfont 62}} & 
\multicolumn{1}{|p{0.23in}}{\Centering {\fontsize{10pt}{12.0pt}\selectfont -95}} & 
\multicolumn{1}{|p{0.26in}}{\Centering {\fontsize{10pt}{12.0pt}\selectfont 31.76}} & 
\multicolumn{1}{|p{0.27in}}{\Centering {\fontsize{10pt}{12.0pt}\selectfont 32.14}} & 
\multicolumn{1}{|p{0.23in}}{\Centering {\fontsize{10pt}{12.0pt}\selectfont 2.40}} & 
\multicolumn{1}{|p{0.23in}}{\Centering {\fontsize{10pt}{12.0pt}\selectfont 2.54}} & 
\multicolumn{1}{|p{0.26in}}{\Centering {\fontsize{10pt}{12.0pt}\selectfont 85.76}} & 
\multicolumn{1}{|p{0.33in}}{\Centering {\fontsize{10pt}{12.0pt}\selectfont 85.88}} & 
\multicolumn{1}{|p{0.23in}|}{\Centering {\fontsize{10pt}{12.0pt}\selectfont 18.0}} \\
\hhline{-------------}
%row no:7
\multicolumn{1}{|p{0.22in}}{\cellcolor[HTML]{FFFFFF}\Centering {\fontsize{10pt}{12.0pt}\selectfont 6}} & 
\multicolumn{1}{|p{0.13in}}{\cellcolor[HTML]{FFFFFF}\Centering {\fontsize{10pt}{12.0pt}\selectfont 3.8}} & 
\multicolumn{1}{|p{0.23in}}{\cellcolor[HTML]{FFFFFF}\Centering {\fontsize{10pt}{12.0pt}\selectfont 31}} & 
\multicolumn{1}{|p{0.23in}}{\cellcolor[HTML]{FFFFFF}\Centering {\fontsize{10pt}{12.0pt}\selectfont -31}} & 
\multicolumn{1}{|p{0.24in}}{\cellcolor[HTML]{FFFFFF}\Centering {\fontsize{10pt}{12.0pt}\selectfont 32}} & 
\multicolumn{1}{|p{0.23in}}{\cellcolor[HTML]{FFFFFF}\Centering {\fontsize{10pt}{12.0pt}\selectfont -25}} & 
\multicolumn{1}{|p{0.26in}}{\cellcolor[HTML]{FFFFFF}\Centering {\fontsize{10pt}{12.0pt}\selectfont 42.30}} & 
\multicolumn{1}{|p{0.27in}}{\cellcolor[HTML]{FFFFFF}\Centering {\fontsize{10pt}{12.0pt}\selectfont 42.99}} & 
\multicolumn{1}{|p{0.23in}}{\cellcolor[HTML]{FFFFFF}\Centering {\fontsize{10pt}{12.0pt}\selectfont 1.70}} & 
\multicolumn{1}{|p{0.23in}}{\cellcolor[HTML]{FFFFFF}\Centering {\fontsize{10pt}{12.0pt}\selectfont 1.84}} & 
\multicolumn{1}{|p{0.26in}}{\cellcolor[HTML]{FFFFFF}\Centering {\fontsize{10pt}{12.0pt}\selectfont 104.9}} & 
\multicolumn{1}{|p{0.33in}}{\cellcolor[HTML]{FFFFFF}\Centering {\fontsize{10pt}{12.0pt}\selectfont 105.18}} & 
\multicolumn{1}{|p{0.23in}|}{\cellcolor[HTML]{FFFFFF}\Centering {\fontsize{10pt}{12.0pt}\selectfont 18.2}} \\
\hhline{-------------}
%row no:8
\multicolumn{1}{|p{0.22in}}{\Centering {\fontsize{10pt}{12.0pt}\selectfont 7}} & 
\multicolumn{1}{|p{0.13in}}{\Centering {\fontsize{10pt}{12.0pt}\selectfont 4.6}} & 
\multicolumn{1}{|p{0.23in}}{\Centering {\fontsize{10pt}{12.0pt}\selectfont 64}} & 
\multicolumn{1}{|p{0.26in}}{\Centering {\fontsize{10pt}{12.0pt}\selectfont -121}} & 
\multicolumn{1}{|p{0.24in}}{\Centering {\fontsize{10pt}{12.0pt}\selectfont 73}} & 
\multicolumn{1}{|p{0.26in}}{\Centering {\fontsize{10pt}{12.0pt}\selectfont -123}} & 
\multicolumn{1}{|p{0.26in}}{\Centering {\fontsize{10pt}{12.0pt}\selectfont 29.10}} & 
\multicolumn{1}{|p{0.27in}}{\Centering {\fontsize{10pt}{12.0pt}\selectfont 29.19}} & 
\multicolumn{1}{|p{0.23in}}{\Centering {\fontsize{10pt}{12.0pt}\selectfont 2.80}} & 
\multicolumn{1}{|p{0.23in}}{\Centering {\fontsize{10pt}{12.0pt}\selectfont 2.83}} & 
\multicolumn{1}{|p{0.26in}}{\Centering {\fontsize{10pt}{12.0pt}\selectfont 80.31}} & 
\multicolumn{1}{|p{0.33in}}{\Centering {\fontsize{10pt}{12.0pt}\selectfont 80.30}} & 
\multicolumn{1}{|p{0.23in}|}{\Centering {\fontsize{10pt}{12.0pt}\selectfont 17}} \\
\hhline{-------------}
%row no:9
\multicolumn{1}{|p{0.22in}}{\cellcolor[HTML]{FFFFFF}\Centering {\fontsize{10pt}{12.0pt}\selectfont 8}} & 
\multicolumn{1}{|p{0.13in}}{\cellcolor[HTML]{FFFFFF}\Centering {\fontsize{10pt}{12.0pt}\selectfont 3.2}} & 
\multicolumn{1}{|p{0.23in}}{\cellcolor[HTML]{FFFFFF}\Centering {\fontsize{10pt}{12.0pt}\selectfont 42}} & 
\multicolumn{1}{|p{0.23in}}{\cellcolor[HTML]{FFFFFF}\Centering {\fontsize{10pt}{12.0pt}\selectfont -46}} & 
\multicolumn{1}{|p{0.24in}}{\cellcolor[HTML]{FFFFFF}\Centering {\fontsize{10pt}{12.0pt}\selectfont 42}} & 
\multicolumn{1}{|p{0.23in}}{\cellcolor[HTML]{FFFFFF}\Centering {\fontsize{10pt}{12.0pt}\selectfont -48}} & 
\multicolumn{1}{|p{0.26in}}{\cellcolor[HTML]{FFFFFF}\Centering {\fontsize{10pt}{12.0pt}\selectfont 39.32}} & 
\multicolumn{1}{|p{0.27in}}{\cellcolor[HTML]{FFFFFF}\Centering {\fontsize{10pt}{12.0pt}\selectfont 38.67}} & 
\multicolumn{1}{|p{0.23in}}{\cellcolor[HTML]{FFFFFF}\Centering {\fontsize{10pt}{12.0pt}\selectfont 2.20}} & 
\multicolumn{1}{|p{0.23in}}{\cellcolor[HTML]{FFFFFF}\Centering {\fontsize{10pt}{12.0pt}\selectfont 2.07}} & 
\multicolumn{1}{|p{0.26in}}{\cellcolor[HTML]{FFFFFF}\Centering {\fontsize{10pt}{12.0pt}\selectfont 98.14}} & 
\multicolumn{1}{|p{0.33in}}{\cellcolor[HTML]{FFFFFF}\Centering {\fontsize{10pt}{12.0pt}\selectfont 97.69}} & 
\multicolumn{1}{|p{0.23in}|}{\cellcolor[HTML]{FFFFFF}\Centering {\fontsize{10pt}{12.0pt}\selectfont 15.6}} \\
\hhline{-------------}

\end{tabular}
 \end{table*}

%%%%%%%%%%%%%%%%%%%% Table No: 7 ends here %%%%%%%%%%%%%%%%%%%%

%%%%%%%%%%%%%%%%%%%% Table No: 8 starts here %%%%%%%%%%%%%%%%%%%%

\begin{table*}
 			\centering
			\caption{Parameters of \textit{c\textsubscript{1}-c\textsubscript{15}},  required in CO$_2$ solubility equations}
\begin{tabular}{p{1.1in}p{1.1in}p{1.1in}p{1.1in}p{1.1in}}
\hline
%row no:1
\multicolumn{1}{|p{1.1in}}{\Centering {\fontsize{10pt}{12.0pt}\selectfont Constant}} & 
\multicolumn{1}{|p{1.1in}}{\Centering  \( \frac{ \mu_\mathrm{CO_2}^{1 \left( 0 \right) }}{RT} \) } & 
\multicolumn{1}{|p{1.1in}}{\Centering  \(  \lambda_\mathrm{CO_2-Na} \) } & 
\multicolumn{1}{|p{1.1in}}{\Centering  \(  \zeta _\mathrm{CO_2-Na-Cl}m_\mathrm{Cl} \) } & 
\multicolumn{1}{|p{1.1in}|}{\Centering  \(  \phi _\mathrm{CO_2} \) } \\
\hhline{-----}
%row no:2
\multicolumn{1}{|p{1.1in}}{\Centering {\fontsize{10pt}{12.0pt}\selectfont $c_1$}} & 
\multicolumn{1}{|p{1.1in}}{\Centering {\fontsize{10pt}{12.0pt}\selectfont 28.9447706}} & 
\multicolumn{1}{|p{1.1in}}{\Centering {\fontsize{10pt}{12.0pt}\selectfont -0.411370585}} & 
\multicolumn{1}{|p{1.1in}}{\Centering {\fontsize{10pt}{12.0pt}\selectfont 3.36389723E-4}} & 
\multicolumn{1}{|p{1.1in}|}{\Centering {\fontsize{10pt}{12.0pt}\selectfont 1.0}} \\
\hhline{-----}
%row no:3
\multicolumn{1}{|p{1.1in}}{\Centering {\fontsize{10pt}{12.0pt}\selectfont $c_2$}} & 
\multicolumn{1}{|p{1.1in}}{\Centering {\fontsize{10pt}{12.0pt}\selectfont -0.0354581768}} & 
\multicolumn{1}{|p{1.1in}}{\Centering {\fontsize{10pt}{12.0pt}\selectfont 6.07632013E-4}} & 
\multicolumn{1}{|p{1.1in}}{\Centering {\fontsize{10pt}{12.0pt}\selectfont -1.98298980E-5}} & 
\multicolumn{1}{|p{1.1in}|}{\Centering {\fontsize{10pt}{12.0pt}\selectfont \textcolor[HTML]{231F20}{4.7586835E-3}}} \\
\hhline{-----}
%row no:4
\multicolumn{1}{|p{1.1in}}{\Centering {\fontsize{10pt}{12.0pt}\selectfont $c_3$}} & 
\multicolumn{1}{|p{1.1in}}{\Centering {\fontsize{10pt}{12.0pt}\selectfont -4770.67077}} & 
\multicolumn{1}{|p{1.1in}}{\Centering {\fontsize{10pt}{12.0pt}\selectfont 97.5347708}} & 
\multicolumn{1}{|p{1.1in}}{\Centering {\fontsize{10pt}{12.0pt}\selectfont -}} & 
\multicolumn{1}{|p{1.1in}|}{\Centering {\fontsize{10pt}{12.0pt}\selectfont \textcolor[HTML]{231F20}{-3.3569963E-6}}} \\
\hhline{-----}
%row no:5
\multicolumn{1}{|p{1.1in}}{\Centering {\fontsize{10pt}{12.0pt}\selectfont $c_4$}} & 
\multicolumn{1}{|p{1.1in}}{\Centering {\fontsize{10pt}{12.0pt}\selectfont 1.02782768E-5}} & 
\multicolumn{1}{|p{1.1in}}{\Centering {\fontsize{10pt}{12.0pt}\selectfont -}} & 
\multicolumn{1}{|p{1.1in}}{\Centering {\fontsize{10pt}{12.0pt}\selectfont -}} & 
\multicolumn{1}{|p{1.1in}|}{\Centering {\fontsize{10pt}{12.0pt}\selectfont 0.0}} \\
\hhline{-----}
%row no:6
\multicolumn{1}{|p{1.1in}}{\Centering {\fontsize{10pt}{12.0pt}\selectfont $c_5$}} & 
\multicolumn{1}{|p{1.1in}}{\Centering {\fontsize{10pt}{12.0pt}\selectfont 33.8126098}} & 
\multicolumn{1}{|p{1.1in}}{\Centering {\fontsize{10pt}{12.0pt}\selectfont -}} & 
\multicolumn{1}{|p{1.1in}}{\Centering {\fontsize{10pt}{12.0pt}\selectfont -}} & 
\multicolumn{1}{|p{1.1in}|}{\Centering {\fontsize{10pt}{12.0pt}\selectfont \textcolor[HTML]{231F20}{-1.3179396}}} \\
\hhline{-----}
%row no:7
\multicolumn{1}{|p{1.1in}}{\Centering {\fontsize{10pt}{12.0pt}\selectfont $c_6$}} & 
\multicolumn{1}{|p{1.1in}}{\Centering {\fontsize{10pt}{12.0pt}\selectfont 9.04037140E-3}} & 
\multicolumn{1}{|p{1.1in}}{\Centering {\fontsize{10pt}{12.0pt}\selectfont -}} & 
\multicolumn{1}{|p{1.1in}}{\Centering {\fontsize{10pt}{12.0pt}\selectfont -}} & 
\multicolumn{1}{|p{1.1in}|}{\Centering {\fontsize{10pt}{12.0pt}\selectfont \textcolor[HTML]{231F20}{-3.8389101E-6}}} \\
\hhline{-----}
%row no:8
\multicolumn{1}{|p{1.1in}}{\Centering {\fontsize{10pt}{12.0pt}\selectfont $c_7$}} & 
\multicolumn{1}{|p{1.1in}}{\Centering {\fontsize{10pt}{12.0pt}\selectfont -1.14934031E-3}} & 
\multicolumn{1}{|p{1.1in}}{\Centering {\fontsize{10pt}{12.0pt}\selectfont -}} & 
\multicolumn{1}{|p{1.1in}}{\Centering {\fontsize{10pt}{12.0pt}\selectfont -}} & 
\multicolumn{1}{|p{1.1in}|}{\Centering {\fontsize{10pt}{12.0pt}\selectfont 0.0}} \\
\hhline{-----}
%row no:9
\multicolumn{1}{|p{1.1in}}{\Centering {\fontsize{10pt}{12.0pt}\selectfont $c_8$}} & 
\multicolumn{1}{|p{1.1in}}{\Centering {\fontsize{10pt}{12.0pt}\selectfont -0.307405726}} & 
\multicolumn{1}{|p{1.1in}}{\Centering {\fontsize{10pt}{12.0pt}\selectfont -0.0237622469}} & 
\multicolumn{1}{|p{1.1in}}{\Centering {\fontsize{10pt}{12.0pt}\selectfont 2.12220830E-3}} & 
\multicolumn{1}{|p{1.1in}|}{\Centering {\fontsize{10pt}{12.0pt}\selectfont \textcolor[HTML]{231F20}{2.2815104E-3}}} \\
\hhline{-----}
%row no:10
\multicolumn{1}{|p{1.1in}}{\Centering {\fontsize{10pt}{12.0pt}\selectfont $c_9$}} & 
\multicolumn{1}{|p{1.1in}}{\Centering {\fontsize{10pt}{12.0pt}\selectfont -0.0907301486}} & 
\multicolumn{1}{|p{1.1in}}{\Centering {\fontsize{10pt}{12.0pt}\selectfont 0.0170656236}} & 
\multicolumn{1}{|p{1.1in}}{\Centering {\fontsize{10pt}{12.0pt}\selectfont -5.24873303E-3}} & 
\multicolumn{1}{|p{1.1in}|}{\Centering {\fontsize{10pt}{12.0pt}\selectfont 0.0}} \\
\hhline{-----}
%row no:11
\multicolumn{1}{|p{1.1in}}{\Centering {\fontsize{10pt}{12.0pt}\selectfont $c_{10}$}} & 
\multicolumn{1}{|p{1.1in}}{\Centering {\fontsize{10pt}{12.0pt}\selectfont 9.32713393E-4}} & 
\multicolumn{1}{|p{1.1in}}{\Centering {\fontsize{10pt}{12.0pt}\selectfont -}} & 
\multicolumn{1}{|p{1.1in}}{\Centering {\fontsize{10pt}{12.0pt}\selectfont -}} & 
\multicolumn{1}{|p{1.1in}|}{\Centering {\fontsize{10pt}{12.0pt}\selectfont 0.0}} \\
\hhline{-----}
%row no:12
\multicolumn{1}{|p{1.1in}}{\Centering {\fontsize{10pt}{12.0pt}\selectfont $c_{11}$}} & 
\multicolumn{1}{|p{1.1in}}{\Centering {\fontsize{10pt}{12.0pt}\selectfont -}} & 
\multicolumn{1}{|p{1.1in}}{\Centering {\fontsize{10pt}{12.0pt}\selectfont 1.41335834E-5}} & 
\multicolumn{1}{|p{1.1in}}{\Centering {\fontsize{10pt}{12.0pt}\selectfont -}} & 
\multicolumn{1}{|p{1.1in}|}{\Centering {\fontsize{10pt}{12.0pt}\selectfont 0.0}} \\
\hhline{-----}

\end{tabular}
 \end{table*}

%%%%%%%%%%%%%%%%%%%% Table No: 8 ends here %%%%%%%%%%%%%%%%%%%%

%%%%%%%%%%%%%%%%%%%% Table No: 9 starts here %%%%%%%%%%%%%%%%%%%%

\begin{table*}
 			\centering
			\caption{Required parameters for the viscosity models}
\begin{tabular}{p{1.96in}p{1.96in}p{1.96in}}
\hline
%row no:1
\multicolumn{1}{|p{1.96in}}{\Centering {\fontsize{10pt}{12.0pt}\selectfont Constant}} & 
\multicolumn{1}{|p{1.96in}}{\Centering {\fontsize{10pt}{12.0pt}\selectfont Mao $\&$  Duan’s model \citep{mao2009viscosity}}} & 
\multicolumn{1}{|p{1.96in}|}{\Centering {\fontsize{10pt}{12.0pt}\selectfont Zhang et al. model \citep{zhang1997viscosity}}} \\
\hhline{---}
%row no:2
\multicolumn{1}{|p{1.96in}}{\Centering {\fontsize{10pt}{12.0pt}\selectfont $c_1$}} & 
\multicolumn{1}{|p{1.96in}}{\Centering {\fontsize{10pt}{12.0pt}\selectfont -0.21319213}} & 
\multicolumn{1}{|p{1.96in}|}{\Centering {\fontsize{10pt}{12.0pt}\selectfont 0.0061}} \\
\hhline{---}
%row no:3
\multicolumn{1}{|p{1.96in}}{\Centering {\fontsize{10pt}{12.0pt}\selectfont $c_2$}} & 
\multicolumn{1}{|p{1.96in}}{\Centering {\fontsize{10pt}{12.0pt}\selectfont 0.13651589E-2}} & 
\multicolumn{1}{|p{1.96in}|}{\Centering {\fontsize{10pt}{12.0pt}\selectfont -}} \\
\hhline{---}
%row no:4
\multicolumn{1}{|p{1.96in}}{\Centering {\fontsize{10pt}{12.0pt}\selectfont $c_3$}} & 
\multicolumn{1}{|p{1.96in}}{\Centering {\fontsize{10pt}{12.0pt}\selectfont -0.12191756E-5}} & 
\multicolumn{1}{|p{1.96in}|}{\Centering {\fontsize{10pt}{12.0pt}\selectfont 0.01040}} \\
\hhline{---}
%row no:5
\multicolumn{1}{|p{1.96in}}{\Centering {\fontsize{10pt}{12.0pt}\selectfont $c_4$}} & 
\multicolumn{1}{|p{1.96in}}{\Centering {\fontsize{10pt}{12.0pt}\selectfont 0.69161945E-1}} & 
\multicolumn{1}{|p{1.96in}|}{\Centering {\fontsize{10pt}{12.0pt}\selectfont 0.000756}} \\
\hhline{---}
%row no:6
\multicolumn{1}{|p{1.96in}}{\Centering {\fontsize{10pt}{12.0pt}\selectfont $c_5$}} & 
\multicolumn{1}{|p{1.96in}}{\Centering {\fontsize{10pt}{12.0pt}\selectfont -0.27292263E-3}} & 
\multicolumn{1}{|p{1.96in}|}{\Centering {\fontsize{10pt}{12.0pt}\selectfont -}} \\
\hhline{---}
%row no:7
\multicolumn{1}{|p{1.96in}}{\Centering {\fontsize{10pt}{12.0pt}\selectfont $c_6$}} & 
\multicolumn{1}{|p{1.96in}}{\Centering {\fontsize{10pt}{12.0pt}\selectfont 0.20852448E-6}} & 
\multicolumn{1}{|p{1.96in}|}{\Centering {\fontsize{10pt}{12.0pt}\selectfont 0.0157}} \\
\hhline{---}
%row no:8
\multicolumn{1}{|p{1.96in}}{\Centering {\fontsize{10pt}{12.0pt}\selectfont $c_7$}} & 
\multicolumn{1}{|p{1.96in}}{\Centering {\fontsize{10pt}{12.0pt}\selectfont -0.25988855E-2}} & 
\multicolumn{1}{|p{1.96in}|}{\Centering {\fontsize{10pt}{12.0pt}\selectfont 0.271}} \\
\hhline{---}
%row no:9
\multicolumn{1}{|p{1.96in}}{\Centering {\fontsize{10pt}{12.0pt}\selectfont $c_8$}} & 
\multicolumn{1}{|p{1.96in}}{\Centering {\fontsize{10pt}{12.0pt}\selectfont 0.77989227E-5}} & 
\multicolumn{1}{|p{1.96in}|}{\Centering {\fontsize{10pt}{12.0pt}\selectfont 0.04712}} \\
\hhline{---}
%row no:10
\multicolumn{1}{|p{1.96in}}{\Centering {\fontsize{10pt}{12.0pt}\selectfont $c_9$}} & 
\multicolumn{1}{|p{1.96in}}{\Centering {\fontsize{10pt}{12.0pt}\selectfont -}} & 
\multicolumn{1}{|p{1.96in}|}{\Centering {\fontsize{10pt}{12.0pt}\selectfont 0.00941}} \\
\hhline{---}
%row no:11
\multicolumn{1}{|p{1.96in}}{\Centering {\fontsize{10pt}{12.0pt}\selectfont $c_{10}$}} & 
\multicolumn{1}{|p{1.96in}}{\Centering {\fontsize{10pt}{12.0pt}\selectfont -}} & 
\multicolumn{1}{|p{1.96in}|}{\Centering {\fontsize{10pt}{12.0pt}\selectfont 0.00003}} \\
\hhline{---}

\end{tabular}
 \end{table*}

%%%%%%%%%%%%%%%%%%%% Table No: 9 ends here %%%%%%%%%%%%%%%%%%%%

%%%%%%%%%%%%%%%%%%%% Table No: 10 starts here %%%%%%%%%%%%%%%%%%%%

\begin{table*}[h]
 			\centering
			\caption{Parameters for the calculation of   NaCl + CaCl$_2$ brine density}
\begin{tabular}{p{0.7in}p{0.64in}p{0.6in}p{0.59in}p{0.59in}p{0.59in}p{0.59in}p{0.59in}}
\hline
%row no:1
\multicolumn{1}{|p{0.7in}}{\cellcolor[HTML]{FFFFFF}\Centering {\fontsize{9pt}{10.8pt}\selectfont Con.}} & 
\multicolumn{1}{|p{0.64in}}{\cellcolor[HTML]{FFFFFF}\Centering {\fontsize{9pt}{10.8pt}\selectfont {$ \alpha_{10}$}}} & 
\multicolumn{1}{|p{0.6in}}{\cellcolor[HTML]{FFFFFF}\Centering {\fontsize{9pt}{10.8pt}\selectfont {$ \alpha_{11}$}}} & 
\multicolumn{1}{|p{0.59in}}{\cellcolor[HTML]{FFFFFF}\Centering {\fontsize{9pt}{10.8pt}\selectfont {$ \alpha_{12}$}}} & 
\multicolumn{1}{|p{0.59in}}{\cellcolor[HTML]{FFFFFF}\Centering {\fontsize{9pt}{10.8pt}\selectfont {$ \alpha_{13}$}}} & 
\multicolumn{1}{|p{0.59in}}{\cellcolor[HTML]{FFFFFF}\Centering {\fontsize{9pt}{10.8pt}\selectfont {$ \alpha_{14}$}}} & 
\multicolumn{1}{|p{0.59in}}{\cellcolor[HTML]{FFFFFF}\Centering {\fontsize{9pt}{10.8pt}\selectfont {$ \alpha_{20}$}}} & 
\multicolumn{1}{|p{0.59in}|}{\cellcolor[HTML]{FFFFFF}\Centering {\fontsize{9pt}{10.8pt}\selectfont {$ \alpha_{21}$}}} \\
\hhline{--------}
%row no:2
\multicolumn{1}{|p{0.7in}}{\Centering {\fontsize{9pt}{10.8pt}\selectfont NaCl}} & 
\multicolumn{1}{|p{0.64in}}{\Centering {\fontsize{9pt}{10.8pt}\selectfont 2863.158}} & 
\multicolumn{1}{|p{0.6in}}{\Centering {\fontsize{9pt}{10.8pt}\selectfont -46844.356}} & 
\multicolumn{1}{|p{0.59in}}{\Centering {\fontsize{9pt}{10.8pt}\selectfont 120760.118}} & 
\multicolumn{1}{|p{0.67in}}{\Centering {\fontsize{9pt}{10.8pt}\selectfont -116867.722}} & 
\multicolumn{1}{|p{0.59in}}{\Centering {\fontsize{9pt}{10.8pt}\selectfont 40285.426}} & 
\multicolumn{1}{|p{0.59in}}{\Centering {\fontsize{9pt}{10.8pt}\selectfont -2000.028}} & 
\multicolumn{1}{|p{0.59in}|}{\Centering {\fontsize{9pt}{10.8pt}\selectfont 34013.518}} \\
\hhline{--------}
%row no:3
\multicolumn{1}{|p{0.7in}}{\Centering {\fontsize{9pt}{10.8pt}\selectfont CaCl$_2$}} & 
\multicolumn{1}{|p{0.64in}}{\Centering {\fontsize{9pt}{10.8pt}\selectfont 2546.760}} & 
\multicolumn{1}{|p{0.6in}}{\Centering {\fontsize{9pt}{10.8pt}\selectfont -39884.946}} & 
\multicolumn{1}{|p{0.59in}}{\Centering {\fontsize{9pt}{10.8pt}\selectfont 102056.957}} & 
\multicolumn{1}{|p{0.67in}}{\Centering {\fontsize{9pt}{10.8pt}\selectfont -98403.334}} & 
\multicolumn{1}{|p{0.59in}}{\Centering {\fontsize{9pt}{10.8pt}\selectfont 33976.048}} & 
\multicolumn{1}{|p{0.59in}}{\Centering {\fontsize{9pt}{10.8pt}\selectfont -1362.157}} & 
\multicolumn{1}{|p{0.59in}|}{\Centering {\fontsize{9pt}{10.8pt}\selectfont 22785.572}} \\
\hhline{--------}
%row no:4
\multicolumn{1}{|p{0.7in}}{\cellcolor[HTML]{FFFFFF}\Centering {\fontsize{9pt}{10.8pt}\selectfont Con.}} & 
\multicolumn{1}{|p{0.64in}}{\cellcolor[HTML]{FFFFFF}\Centering {\fontsize{9pt}{10.8pt}\selectfont {$ \alpha_{22}$}}} & 
\multicolumn{1}{|p{0.6in}}{\cellcolor[HTML]{FFFFFF}\Centering {\fontsize{9pt}{10.8pt}\selectfont {$ \alpha_{23}$}}} & 
\multicolumn{1}{|p{0.59in}}{\cellcolor[HTML]{FFFFFF}\Centering {\fontsize{9pt}{10.8pt}\selectfont {$ \alpha_{24}$}}} & 
\multicolumn{1}{|p{0.59in}}{\cellcolor[HTML]{FFFFFF}\Centering {\fontsize{9pt}{10.8pt}\selectfont {$ \alpha_{30}$}}} & 
\multicolumn{1}{|p{0.59in}}{\cellcolor[HTML]{FFFFFF}\Centering {\fontsize{9pt}{10.8pt}\selectfont {$ \alpha_{31}$}}} & 
\multicolumn{1}{|p{0.59in}}{\cellcolor[HTML]{FFFFFF}\Centering {\fontsize{9pt}{10.8pt}\selectfont {$ \alpha_{32}$}}} & 
\multicolumn{1}{|p{0.59in}|}{\cellcolor[HTML]{FFFFFF}\Centering {\fontsize{9pt}{10.8pt}\selectfont {$ \alpha_{33}$}}} \\
\hhline{--------}
%row no:5
\multicolumn{1}{|p{0.7in}}{\Centering {\fontsize{9pt}{10.8pt}\selectfont NaCl}} & 
\multicolumn{1}{|p{0.64in}}{\Centering {\fontsize{9pt}{10.8pt}\selectfont -88557.123}} & 
\multicolumn{1}{|p{0.6in}}{\Centering {\fontsize{9pt}{10.8pt}\selectfont 86351.784}} & 
\multicolumn{1}{|p{0.67in}}{\Centering {\fontsize{9pt}{10.8pt}\selectfont -29910.216}} & 
\multicolumn{1}{|p{0.59in}}{\Centering {\fontsize{9pt}{10.8pt}\selectfont 413.046}} & 
\multicolumn{1}{|p{0.59in}}{\Centering {\fontsize{9pt}{10.8pt}\selectfont -7125.857}} & 
\multicolumn{1}{|p{0.59in}}{\Centering {\fontsize{9pt}{10.8pt}\selectfont 18640.780}} & 
\multicolumn{1}{|p{0.67in}|}{\Centering {\fontsize{9pt}{10.8pt}\selectfont -18244.074}} \\
\hhline{--------}
%row no:6
\multicolumn{1}{|p{0.7in}}{\Centering {\fontsize{9pt}{10.8pt}\selectfont CaCl$_2$}} & 
\multicolumn{1}{|p{0.64in}}{\Centering {\fontsize{9pt}{10.8pt}\selectfont -59216.108}} & 
\multicolumn{1}{|p{0.6in}}{\Centering {\fontsize{9pt}{10.8pt}\selectfont 57894.824}} & 
\multicolumn{1}{|p{0.67in}}{\Centering {\fontsize{9pt}{10.8pt}\selectfont -20222.898}} & 
\multicolumn{1}{|p{0.59in}}{\Centering {\fontsize{9pt}{10.8pt}\selectfont 217.778}} & 
\multicolumn{1}{|p{0.59in}}{\Centering {\fontsize{9pt}{10.8pt}\selectfont -3770.645}} & 
\multicolumn{1}{|p{0.59in}}{\Centering {\fontsize{9pt}{10.8pt}\selectfont 9908.135}} & 
\multicolumn{1}{|p{0.59in}|}{\Centering {\fontsize{9pt}{10.8pt}\selectfont -9793.484}} \\
\hhline{--------}
%row no:7
\multicolumn{1}{|p{0.7in}}{\cellcolor[HTML]{FFFFFF}\Centering {\fontsize{9pt}{10.8pt}\selectfont Con.}} & 
\multicolumn{1}{|p{0.64in}}{\cellcolor[HTML]{FFFFFF}\Centering {\fontsize{9pt}{10.8pt}\selectfont {$ \alpha_{34}$}}} & 
\multicolumn{1}{|p{0.6in}}{\cellcolor[HTML]{FFFFFF}\Centering {\fontsize{9pt}{10.8pt}\selectfont {$ \beta_{10}$}}} & 
\multicolumn{1}{|p{0.59in}}{\cellcolor[HTML]{FFFFFF}\Centering {\fontsize{9pt}{10.8pt}\selectfont {$ \beta_{11}$}}} & 
\multicolumn{1}{|p{0.59in}}{\cellcolor[HTML]{FFFFFF}\Centering {\fontsize{9pt}{10.8pt}\selectfont {$ \beta_{12}$}}} & 
\multicolumn{1}{|p{0.59in}}{\cellcolor[HTML]{FFFFFF}\Centering {\fontsize{9pt}{10.8pt}\selectfont {$ \beta_{13}$}}} & 
\multicolumn{1}{|p{0.59in}}{\cellcolor[HTML]{FFFFFF}\Centering {\fontsize{9pt}{10.8pt}\selectfont {$ \gamma_{1}$}}} & 
\multicolumn{1}{|p{0.59in}|}{\cellcolor[HTML]{FFFFFF}\Centering {\fontsize{9pt}{10.8pt}\selectfont {$ \gamma_{2}$}}} \\
\hhline{--------}
%row no:8
\multicolumn{1}{|p{0.7in}}{\Centering {\fontsize{9pt}{10.8pt}\selectfont NaCl}} & 
\multicolumn{1}{|p{0.64in}}{\Centering {\fontsize{9pt}{10.8pt}\selectfont 6335.275}} & 
\multicolumn{1}{|p{0.6in}}{\Centering {\fontsize{9pt}{10.8pt}\selectfont 241.57}} & 
\multicolumn{1}{|p{0.59in}}{\Centering {\fontsize{9pt}{10.8pt}\selectfont -980.97}} & 
\multicolumn{1}{|p{0.59in}}{\Centering {\fontsize{9pt}{10.8pt}\selectfont 1482.31}} & 
\multicolumn{1}{|p{0.59in}}{\Centering {\fontsize{9pt}{10.8pt}\selectfont -750.98}} & 
\multicolumn{1}{|p{0.59in}}{\Centering {\fontsize{9pt}{10.8pt}\selectfont -0.00134}} & 
\multicolumn{1}{|p{0.59in}|}{\Centering {\fontsize{9pt}{10.8pt}\selectfont 0.00056}} \\
\hhline{--------}
%row no:9
\multicolumn{1}{|p{0.7in}}{\Centering {\fontsize{9pt}{10.8pt}\selectfont CaCl$_2$}} & 
\multicolumn{1}{|p{0.64in}}{\Centering {\fontsize{9pt}{10.8pt}\selectfont 3455.587}} & 
\multicolumn{1}{|p{0.6in}}{\Centering {\fontsize{9pt}{10.8pt}\selectfont 307.24}} & 
\multicolumn{1}{|p{0.59in}}{\Centering {\fontsize{9pt}{10.8pt}\selectfont -1259.10}} & 
\multicolumn{1}{|p{0.59in}}{\Centering {\fontsize{9pt}{10.8pt}\selectfont 2034.03}} & 
\multicolumn{1}{|p{0.59in}}{\Centering {\fontsize{9pt}{10.8pt}\selectfont -1084.94}} & 
\multicolumn{1}{|p{0.59in}}{\Centering {\fontsize{9pt}{10.8pt}\selectfont -0.00493}} & 
\multicolumn{1}{|p{0.59in}|}{\Centering {\fontsize{9pt}{10.8pt}\selectfont 0.00231}} \\
\hhline{--------}

\end{tabular}
 \end{table*}

%%%%%%%%%%%%%%%%%%%% Table No: 10 ends here %%%%%%%%%%%%%%%%%%%%

\end{document}